\documentclass[11pt]{emulateapj}
\usepackage{amsmath}
\usepackage{amssymb}
\usepackage{natbib}
\usepackage{graphicx}
\newcommand{\D}{\displaystyle}\newcommand{\R}{\scriptstyle}

\def\Wm2{W/m$^2$}
\def\Wpm2sr{Wm$^{-2}sr^{-1}$}
\def\deg{$^\circ$ }
\def\degx{$^\circ$}

\def\etal{{\it et al.\ }}

\begin{document}

\title{Accurate and Approximate Calculations of Raman Scattering in the Atmosphere of Neptune}
 \author{L.A. Sromovsky\altaffilmark{1}}
\altaffiltext{1}{University of Wisconsin - Madison, Madison WI 53706}

\slugcomment{Journal reference: Icarus 173 (2005) 254-283.}

\begin{abstract}
Raman scattering by H$_2$ in Neptune's atmosphere has significant
effects on its reflectivity for $\lambda <$ 0.5 $\mu$m, producing
baseline decreases of $\sim$ 20\% in a clear atmosphere and $\sim$
10\% in a hazy atmosphere. However, few accurate Raman calculations
are carried out because of their complexity and computational costs.
Here we present the first radiation transfer algorithm that includes
both polarization and Raman scattering and facilitates computation of
spatially resolved spectra. New calculations show that Cochran and
Trafton's (1978, {\it Astrophys. J. \bf 219}, 756-762) suggestion that
light reflected in the deep CH$_4$ bands is mainly Raman scattered is
not valid for current estimates of the CH$_4$vertical distribution,
which implies only a 4\% Raman contribution. Comparisons with IUE,
HST, and groundbased observations confirm that high altitude haze
absorption is reducing Neptune's geometric albedo by $\sim$6\% in the
0.22-0.26 $\mu$m range and by $\sim$13\% in the 0.35-0.45 $\mu$m
range. A sample haze model with 0.2 optical depths of 0.2-$\mu$m
radius particles between 0.1 and 0.8 bars fits reasonably well, but is
not a unique solution. We used accurate calculations to evaluate
several approximations of Raman scattering.  The Karkoschka (1994,
{\it Icarus \bf 111}, 174-192) method of applying Raman corrections to
calculated spectra and removing Raman effects from observed spectra is
shown to have limited applicability and to undercorrect the depths of
weak CH$_4$ absorption bands.  The relatively large Q-branch
contribution observed by Karkoschka is shown to be consistent with
current estimates of Raman cross sections.  The Wallace (1972, {\it
  Astrophys. J. \bf 176}, 249-257) approximation, produces geometric
albedo values $\sim$5\% low as originally proposed, but can be made
much more accurate by including a scattering contribution from the
vibrational transition. The original Pollack \etal (1986, {\it Icarus
  \bf 65}, 442-466) approximation is inaccurate and unstable, but can
be greatly improved by several simple modifications.  A new
approximation based on spectral tuning of the effective molecular
single scattering albedo provides low errors for zenith angles below
70\degx in a clear atmosphere, although intermediate clouds present
problems at longer wavelengths.

\end{abstract}

\maketitle
\shortauthors{Sromovsky} 
\shorttitle{Accurate and Approximate Calculations of Raman Scattering.}

\section{Introduction}

Because Neptune's atmosphere has a relatively low burden of aerosols,
its reflected spectrum is strongly influenced by both Rayleigh
scattering and Raman scattering by molecular hydrogen. Rayleigh
scattering induces polarization that can significantly modify the
reflected intensity (Mishchenko \etal 1994), accurate computation of
which presents the very large burden of solving the vector radiation
transfer equation.  Sromovsky (2004) discusses that problem and a new approximation
method applicable to low phase angles.
Accurate treatment of Raman scattering is also a computational burden
because photons incident at one wavelength lose some energy to
rotating and/or vibrating the hydrogen molecule and reappear at longer
wavelengths.  Computation of reflectivity at one wavelength thus
requires accounting for contributions from Raman scattering at shorter
wavelengths. In addition, because the Raman source function varies continuously
with optical depth, otherwise homogeneous layers become
inhomogeneous, requiring many more layers to achieve an accurate
characterization of the atmosphere.  

To avoid the computational burden of rigorous Raman scattering
calculations, several different approximations have been employed.
Baines and Smith (1994) used an approximation suggested by Wallace (1972), in
which the Raman cross sections for rotational transitions are treated
as conservative scattering because the wavelength shifts are
relatively small, while the cross section for the vibrational
transition, which involves a much larger wavelength shift, is treated
as an absorption.  This approximation does not produce the sharp
spectral features characteristic of Raman scattering and is of
uncertain accuracy.
Pollack \etal (1986) used an alternate approximation in which
the Rayleigh scattering cross section at a given wavelength is scaled
by the solar irradiance ratio at the shifted and unshifted
wavelengths. This approximation does produce Raman spectral features,
but the accuracy is not well known, and can create conservation
problems by allowing single-scattering albedo values exceeding unity.
Karkoschka (1994) presented a method for correcting observations to
remove Raman scattering and for converting calculations that ignored
Raman scattering to spectra that approximately matched spectra that
included Raman scattering. That method can add or remove Raman
spectral features and was applied to observations of Saturn, Jupiter,
Uranus, and Neptune, but was never tested for accuracy or generality.

A number of calculations of Neptune's geometric albedo have been made
that do account for the basic physics of Raman scattering.  Cochran
and Trafton (1978) implemented an iterative algorithm in which Raman
scattering is first treated as an absorption. After solving the scalar
radiation transfer equation at each frequency grid point, the photon
loss is computed from the radiation field.  On the next iteration the
lost photons are added back as source terms at the shifted wavelengths
appropriate to each Raman transition.  They achieved convergence after
three iterations. Their mainly low resolution results are of limited
utility however, because the atmospheric structure they assumed is so
different from our current understanding. Their claim that the
residual intensity in the cores of the strong methane bands could be
entirely explained by Raman scattering will be shown to be invalid
because of their assumed CH$_4$ mixing ratio
profile. The first model calculations displaying extensively detailed
Raman spectral features in Neptune's atmosphere are those of Courtin
(1999), who made use of the two-stream code of Toon \etal (1989) to
speed the solution of the radiative transfer equation. But this method
can deviate from exact solutions by 10-15\%, does not account for
polarization, and is not usable for studying center-to-limb
variations.  A more rigorous method was used by B\'{e}tremieux and
Yelle (1999), based on the DISORT radiative transfer code (Stamnes
\etal 1988).  But they presented only results for Jupiter and omitted
polarization effects.

This paper presents a new method for accurate computation of Raman
scattering that includes polarization in the context of Neptune's
atmosphere and makes preliminary applications of that method to
resolve several significant issues. The next section reviews the basic
physics of Raman scattering. That is followed by a discussion of
methods for accurate computation of Raman scattering. Sample
computations are then presented to characterize the basic features of
Raman scattering on Neptune.  Comparisons are made with various prior
calculations, and with HST and groundbased observations, to assess the
degree of haze absorption required in Neptune's atmosphere.  The final
section evaluates past approximations, discusses how they can be
generalized and improved, and presents the new approximation and its
performance.

\section{The Physics of Raman Scattering}

\subsection{Hydrogen Energy Levels and Transitions.}

The hydrogen molecule can exist in two nuclear spin states.  Ortho
states have parallel nuclear spins and odd total angular momentum
quantum numbers ($J=1,3,$...) with a degeneracy of $3(2J+1)$. The para states
have antiparallel nuclear spins and even angular momentum quantum numbers
($J=0,2,$..) with a degeneracy of $(2J+1)$.  The equilibrium population of
these states follows the Boltzmann distribution, so that the fraction
of molecules with angular momentum $J$ is given by 
\begin{equation}
   P_{EQ}(J) = d(J) \exp (-E(J)/kT) / \sum_{J=0}^{\infty}(d(J)\exp(-E(J)/kT))
\end{equation}
where $d(J)$ is the degeneracy, $E(J)$ is the energy above the ground
state, $k$ is the Boltzmann constant, and $T$ is absolute
temperature.  Using published expressions for the energy levels (Farkas
1935; Massie and Hunten 1982) we obtain the fractional populations
given in Table 1, where $f_{para}$ is the total
fraction of molecules with even $J$ and $f_{ortho}$ is the total
fraction of molecules with odd $J$.  This assumes that ortho and para
states can exchange energy. However, the time scale for equilibration
by means of bimolecular collisions is of the order of years (Massie
and Hunten 1982) and radiative energy exchange is
forbidden by selection rules.  Thus it is also meaningful to consider
a different kind of equilibrium condition in which ortho states equilibrate
separately from para states.  This is relevant when hydrogen equilibrates
at high temperature, then is lifted by convection to higher altitude where
the fraction of para and ortho molecules remain fixed (over short time scales)
but equilibration of energy levels within each sub-population does take place.
For ``normal'' H$_2$, which is defined by the high temperature equilibrium value of
$f_{para}=0.25$, the sub-population distributions are given by
 \begin{eqnarray}
 P_{norm}(J) =
\begin{cases}
   \frac{1}{4} d(J) \exp (-E(J)/kT) / Z_P,  &  \text{$J$ even}\\
   \frac{3}{4} d(J) \exp (-E(J)/kT) / Z_O,  & \text{$J$ odd}
\end{cases}\label{Eq:pnorm}
\end{eqnarray}
where the partition functions $Z_P$ and $Z_O$ are summations of $d(J)
\exp (-E(J)/kT)$ carried out over even $J$ and odd $J$ respectively.
This distribution is given in Table\ 2 as a function of temperature. 

\begin{table}
\caption{Fractional Populations of H$_2$ Rotational States for Equilibrium H$_2$.}
\begin{tabular}{c c c c c c c c}
\hline
\rule{0pt}{0.15in}
 T(K)  & J=0   &  J=1   &  J=2  &   J=3 &  J=4 & f$_{p}$   &  f$_{o}$\\
\hline
\rule{0pt}{0.15in}
 50 &  0.7704&  0.2294&  0.0001&  0.0000&  0.0000&  0.771&  0.229\\
 75 &  0.5173&  0.4798&  0.0029&  0.0000&  0.0000&  0.520&  0.480\\
100 &  0.3747&  0.6135&  0.0115&  0.0003&  0.0000&  0.386&  0.614\\
125 &  0.2947&  0.6784&  0.0250&  0.0018&  0.0000&  0.320&  0.680\\
150 &  0.2450&  0.7080&  0.0410&  0.0059&  0.0000&  0.286&  0.714\\
175 &  0.2112&  0.7178&  0.0574&  0.0135&  0.0001&  0.269&  0.731\\
200 &  0.1865&  0.7157&  0.0729&  0.0245&  0.0004&  0.260&  0.740\\
225 &  0.1673&  0.7061&  0.0869&  0.0387&  0.0009&  0.255&  0.745\\
250 &  0.1520&  0.6919&  0.0990&  0.0552&  0.0017&  0.253&  0.747\\
275 &  0.1394&  0.6749&  0.1092&  0.0732&  0.0028&  0.251&  0.749\\
300 &  0.1287&  0.6564&  0.1177&  0.0919&  0.0043&  0.251&  0.749\\[0.5ex]
\hline
\end{tabular}
Note: f$_p$ and f$_o$ denote  f$_{para}$ and $f_{ortho}$ respectively.
\label{Tbl:eqpop}
\end{table}

\begin{table}\centering
\caption{Fractional Populations of H$_2$ Rotational States for Normal H$_2$.}
\begin{tabular}{c c c c c c c c}
\hline
\rule{0pt}{0.15in}
 T(K)  & J=0   &  J=1   &  J=2  &   J=3 &  J=4 & f$_{p}$   &  f$_{o}$\\
 \hline
\rule{0pt}{0.15in}
50 &  0.2500&  0.7500&  0.0000&  0.0000&  0.0000&  0.25 &  0.75\\
 75 &  0.2486&  0.7500&  0.0014&  0.0000&  0.0000&  0.25 &  0.75\\
100 &  0.2426&  0.7496&  0.0074&  0.0004&  0.0000&  0.25 &  0.75\\
125 &  0.2305&  0.7480&  0.0195&  0.0020&  0.0000&  0.25 &  0.75\\
150 &  0.2142&  0.7438&  0.0358&  0.0062&  0.0000&  0.25 &  0.75\\
175 &  0.1965&  0.7362&  0.0534&  0.0138&  0.0001&  0.25 &  0.75\\
200 &  0.1795&  0.7251&  0.0702&  0.0249&  0.0004&  0.25 &  0.75\\
225 &  0.1640&  0.7109&  0.0851&  0.0390&  0.0008&  0.25 &  0.75\\
250 &  0.1504&  0.6944&  0.0979&  0.0554&  0.0016&  0.25 &  0.75\\
275 &  0.1386&  0.6762&  0.1086&  0.0733&  0.0028&  0.25 &  0.75\\
300 &  0.1283&  0.6570&  0.1174&  0.0920&  0.0043&  0.25 &  0.75\\
\hline
\end{tabular}
Note: f$_p$ and f$_o$ denote  f$_{para}$ and $f_{ortho}$ respectively.
\label{Tbl:normpop}
\end{table}

\begin{figure}[!htb]
\hspace{-0.2in}\includegraphics[width=3.7in]{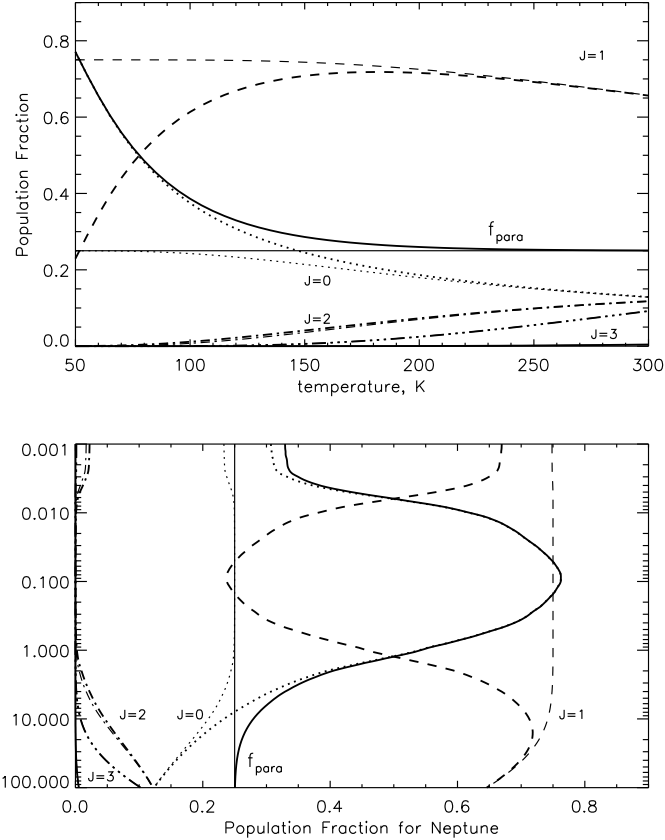}
\caption{TOP: Fractional populations of angular momentum states as
 a function of temperature for equilibrium H$_2$ (thick lines) and for
 normal H$_2$ (thin lines). BOTTOM: Population fractions vs pressure
 in Neptune's atmosphere.}
\label{Fig:parapop}
\end{figure}

The populations for both equilibrium and normal hydrogen are plotted
in Fig. \ref{Fig:parapop}, both as a function of temperature and as a
function of pressure in Neptune's atmosphere.  At T $>$
300 K, $f_{para}$ approaches 0.25 and the ortho/para ratio
approaches the 3/1 ratio expected  from the
nuclear spin degeneracy.  But at the low temperatures in the
upper troposphere and stratosphere of Neptune $f_{para}$ can be
much larger, and for pressures less than a few bars, only $J=0$ and
$J=1$ ground states need to be considered.

Radiative transitions must satisfy selection rules $\Delta J=-2,0,+2$
(Hollas 1992); the corresponding transitions are named $O(J)$, $Q(J)$,
and $S(J)$, where $J$ is the angular momentum quantum number of the
initial state.  The $O$ and $S$ branches can involve changes in
rotational energy and vibrational energy, while the $Q$ branch
involves changes in only vibrational energy. For the $O$ branch
interactions the scattered photons have more energy than the incident
photons. But for the $O$ branch to play a significant role, there
needs to be a significant population in rotational states with $J \geq
2$, which is not the case for the upper troposphere of Neptune.
Thus, we here ignore the $O$ branch.

\subsection{Raman Scattering Cross Sections}

\begin{figure}[!htb]
\hspace{-0.2in}\includegraphics[width=3.7in]{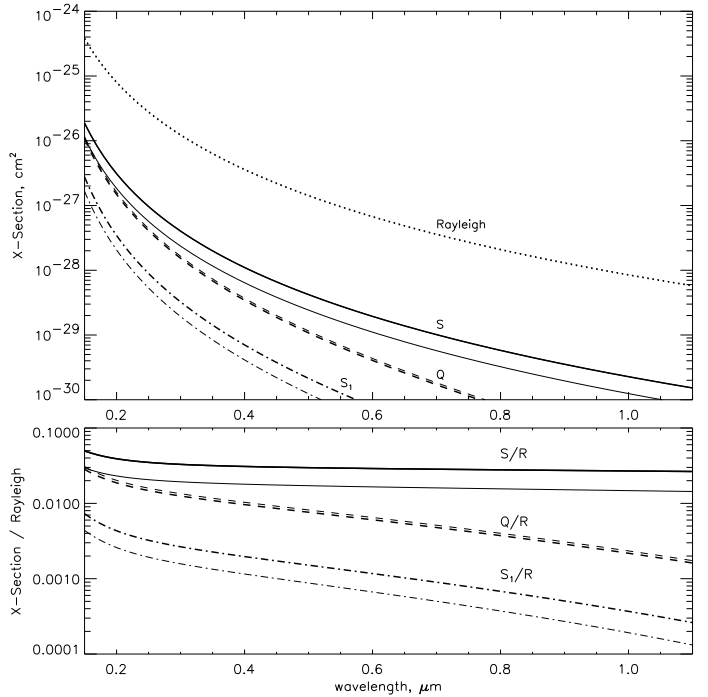}
\caption{TOP: Raman cross sections/molecule vs wavelength for $J=0$ ground states
 (thick lines) and for  $J=1$ ground states (thin
lines), where $S_1$ refers to the $\Delta J=2$, $\Delta v=1$ vibrational
transition. BOTTOM: Ratio of Raman to Rayleigh cross sections.}
\label{Fig:xcvswlen}
\end{figure}

The wavelength-dependent Raman and Rayleigh cross sections per molecule
(Fig.\ \ref{Fig:xcvswlen}) are computed using fits
given by Ford and Browne (1973). Cross sections at 0.4 $\mu$m are
given in Table\ 3, along with transition energy energies expressed
as corresponding wavenumber shifts.  These cross sections for the
$S$ and $Q$ Raman transitions are 46-68\% larger than those given by Cochran
and Trafton (1978).

\begin{table}\centering
\caption{Raman transitions, wavenumber shifts, and 0.4-$\mu$m cross sections 
per molecule.  }
\begin{tabular}{c c c c c}
\hline
\rule{0pt}{0.15in}
  &  &  &  &   0.4-$\mu$m Cross \\
  Transition($J$) & $\Delta J$& $\Delta v$ & $\Delta \nu$, cm$^{-1}$ & section, cm$^{2}$\\
 \hline
 $R(0)$  & 0&0  & 354.69  &  3.575$\times$10$^{-27}$ \\
 $R(1)$  & 0&0  & 587.07  &  3.635$\times$10$^{-27}$ \\
 $S(0)$  & +2&0  & 354.69  & 1.104$\times$10$^{-28}$ \\
 $S(1)$  & +2&0  & 587.07  & 0.642$\times$10$^{-28}$ \\
 $Q(0)$  & 0&1   & 4162.06 &  0.344$\times$10$^{-28}$ \\
 $Q(1)$  & 0&1   & 4156.15 &  0.369$\times$10$^{-28}$ \\
 $S_1(0)$ & +2&1  & 4498.75 &  0.070$\times$10$^{-28}$ \\
 $S_1(1)$ & +2&1  & 4713.83 &  0.041$\times$10$^{-28}$ \\
 $Q$ & &   & 4161.00 &  0.412$\times$10$^{-28}$ \\[0.5ex]
\hline
\end{tabular}
\parbox{3.in}{NOTES: The $R(J)$ transitions refer to Rayleigh scattering
cross sections, which do differ slightly with ground state quantum number.
The subscript in $S_1(J)$ refers to the vibrational quantum number change.
The unsubscripted $Q$ refers to the sum of the four prior cross-sections
divided by two (the last term in Eq.\ \ref{Eq:sigtot}).}
\label{Tbl:xctable}
\end{table}

Note in Table\ 3 that the $Q(0)+S_1(0)$ cross section sum is only
1\% larger than $Q(1)+S_1(1)$ sum at 0.4 $\mu$m.  Fig.\
\ref{Fig:xcvswlen} shows that this close match holds true at all
wavelengths of interest.  These transitions also all have similar
wavenumber shifts. In fact, the shifts of $Q(0)$ and $Q(1)$, which
have the largest cross sections, differ by only 6 wavenumbers, which
is several times smaller than the spectral resolution that we need to
model existing observations.  For these transitions we can thus follow
Courtin (1999) by ignoring $f_{para}$ and simply taking the effective
cross section as the average of $J=0$ and $J=1$ transitions and using a
single wavenumber shift of $4161$ cm$^{-1}$.  
The inclusion of the $S_1$
terms correctly accounts for the absorption of photons at the incident
wavelength, but does not transfer the scattered photons to exactly the
correct longer wavelength.  However, this results in a negligible
error.

The effective scattering cross section per molecule for each
transition depends on the fractional population of the initial state
as well as the transition cross section.  For Neptune's the upper troposphere
 we can make the approximation that there are only two
ground states: $J=0$ and $J=1$.  This means that $f_{para}$ can be
taken as the fraction of $J=0$ molecules, and $f_{ortho}$ as the
fraction of $J=1$ molecules.
We can then write the total Raman scattering
cross section as \begin{eqnarray} \sigma_{total} = f_{para}\D \sigma
\R (S(0)) +
\D (1-f_{para}) \D {\sigma} \R (S(1))\D +\nonumber \\ 
\frac{1}{2}[ \sigma \R (Q(0))\D + \sigma \R (Q(1)) +
\D \sigma \R (S_1(0)) +\D \sigma \R (S_1(1)) \D]
\label{Eq:sigtot}\end{eqnarray}
where we will hereafter refer to the last term as the $Q$ cross
section, as if it were due to a single transition, and its
four individual contributions will be assigned the same
wavenumber shift of 4161 cm$^{1}$.

\subsection{Raman phase functions}\label{Sec:ramphase}

The generalized scalar phase function for scattering of unpolarized
incident radiation by anisotropic molecules for all types of molecular
scattering (including Raman and Rayleigh scattering) can be written as
(Placzek 1959; Soris and Evans 1999):
\begin{equation} P(\theta) = \frac{3}{2}\Big[ \frac{(1+\delta) +
(1-\delta) \cos^2\theta} {2+\delta} \Big]\label{Eq:phase1}
,\end{equation}
where $\delta$ is the depolarization ratio, which is the ratio of
perpendicular to parallel intensities scattered at 90\degx. The
depolarization ratio is zero for isotropic particles (the pure
Rayleigh case) and unity for isotropic scatterers.  For pure
rotational Raman scattering, $\delta=6/7$ (Soris and Evans 1999),
yielding $P(\theta)=\frac{3}{40}[13+\cos^2\theta]$, which has a
peak deviation of only 7.7\% from isotropy.
Depolarization ratios for vibrational transitions 
for gases seem to be generally below
about 0.38 (Bhagavantam 1942). For the 4156 cm$^{-1}$
$Q$-branch transition, measured values range from 0.045 (Cabannes and Rousset
1936) to 0.13 (Bhagavantam 1931). Either value implies a
phase function that is closer to pure Rayleigh than to isotropic.
However, the deviation from isotropic is greatest for the first
scattering, for which the direct beam plays the
largest role.  Backscattered $Q$-branch light from the first
scattering would be reduced by 30-33\% using the assumption of
isotropic scattering.  But in sample calculations the reflected flux
of Raman scattered photons attributable to this transition is only
17\% of the total (see Fig.\ \ref{Fig:ramline1plus2}) and about half of
that comes from incident light that already been diffusely scattered.
Thus, the net effect of assuming isotropic scattering for
the vibrational transition will probably never exceed 15\% of the
$Q$-branch contribution to single scattering, and will generally be
insignificant. Thus we will assume isotropic scattering for both
rotational and vibrational transitions to benefit from the
simplifications that it permits.

\section{Radiation Transfer Methods}

\subsection{Equation of Transfer and Definition of Coordinate System}

We follow Evans and Stephens (1991) and others in defining downward as
positive, so that the direction of light propagation is defined by the
vector ($\theta$, $\phi$) or ($\mu$,$\phi$), where $\theta$ is the
angle from the inward normal at the top of the atmosphere and $\mu =
\cos\theta$.  The angle $\phi$ is the azimuth angle measured
clockwise looking upward.  We assume an unpolarized collimated
incident solar beam in the direction ($\mu_0$, $\phi_0 =0$).  Light
that is scattered backward toward the source then has the direction
($-\mu_0$, $\pi$). The Stokes vector $\vec{\mathsf{I}}$ is defined as a 
4-element column vector given by \begin{equation}
\vec{\mathsf{I}} = \left[ \begin{array}{c} I\\ Q\\ U\\ V\end{array}
 \right]\label{Eq:stokesvec}
\end{equation}
where $I$ is the total intensity and $(Q^2+U^2+V^2)^\frac{1}{2}$ is the intensity of
polarized light. Separate definitions for $Q$, $U$, and $V$ are given by
Hansen and Travis (1974).

With these definitions, the vector radiation transfer equation can be
written as \begin{eqnarray}
&&\mu \frac{d\vec{\mathsf{I}}(z,\mu,\phi)}{dz} = -k_\mathrm{ext}\vec{\mathsf{I}}(z,\mu,\phi) + \nonumber \\
& &k_\mathrm{scatt}\frac{1}{4\pi}
\int^{2\pi}_{0}\int^1_{-1} \mathbf{M}(\mu,\phi,\mu',\phi')
\vec{\mathsf{I}}(z,\mu',\phi') d\mu' d\phi' + \nonumber \\
& &\vec{\mathsf{J}}(z,\mu,\phi) \label{Eq:RTEZ},
\end{eqnarray}
where $z$ denotes distance downward into the atmosphere, $k_\mathrm{ext}$ 
and $k_\mathrm{scatt}$ are linear extinction and scattering coefficients, 
and where $\vec{\mathsf{J}}$ is the source vector due to Raman scattering.  
Equation\ \ref{Eq:RTEZ} is often
written in terms of optical depth $\tau$, where $d\tau =k_\mathrm{ext}dz$, 
and single-scattering albedo
$\omega = k_\mathrm{scatt}/k_\mathrm{ext}$.  However, because Raman
scattering transfers photons from one wavelength to another at a
specific physical location in the atmosphere, the wavelength-independent 
distance $z$ is a more
convenient vertical coordinate than $\tau$.  In Eq.\
\ref{Eq:RTEZ} the scattering matrix $\mathbf{M}$ is a rotated form of
the scattering phase matrix $\mathbf{P}$, given by
\begin{equation}
\mathbf{M}(\mu,\phi,\mu',\phi')=\mathbf{L}(i_2-\pi)\mathbf{P}(\cos\alpha)
\mathbf{L}(i_1) \label{Eq:scatmat}
\end{equation}
where $\alpha$ is the scattering angle, $i_1$ is the angle between the
scattering plane and the meridional plane of the incoming ray, and
$i_2$ is the angle between the scattering plane and the meridional
plane of the outgoing ray.  Formulas for the scattering and rotation
angles, and the form of the rotation matrix, are given by Hansen and
Travis (1974).  In this formulation, the Stokes vector
$\vec{\mathsf{I}}$ includes direct beam as well as diffuse components
of the radiation field, and thus there is no solar pseudo source as
used by Evans and Stephens (1991). There is an implicit wavelength
dependence in Eq.\ \ref{Eq:RTEZ}, but the wavelength dependence must
be made explicit in the equation for the source
vector:\begin{eqnarray} &&\vec{\mathsf{J}}(z,\mu,\phi,\nu)
\frac{1}{4\pi}\sum_{\ell=0}^{n_R} f_\ell n_{H_2}(z)\sigma_{R\ell}(\nu^*_\ell)\nonumber \\
&& \int  \mathbf{M_{R\ell}}(\mu,\phi,\mu',\phi')
)\vec{\mathsf{I}}(z,\mu',\phi',\nu^*_\ell) d\Omega',
\label{Eq:source}\end{eqnarray}
where we use wavenumber $\nu = 1/\lambda$ instead of wavelength
because 
each Raman transition is associated with a fixed shift in wavenumber
but a variable shift in wavelength.  Here the quantity $ n_{H_2}(z)$
is the volume number density of H$_2$ molecules, and
$\sigma_{R\ell}(\nu^*_\ell)$ is the cross section for photons at
wavenumber $\nu^*_\ell = \nu + \Delta \nu_\ell$ to excite transition
$\ell$ and exit at wavenumber $\nu$, where $h\Delta\nu_\ell/c$ is the
energy change associated with the transition. The associated
scattering matrix is $\mathbf{M_{R\ell}}$.  Note that the vector
intensity $\vec{\mathsf{I}}(z,\mu',\phi',\nu^*_\ell)$ in the integrand 
is evaluated at the
wavenumber of the incident photon, as are the cross section and
scattering matrix in Eq.\ \ref{Eq:source}. If $\vec{\mathsf{I}}$
is measured in photons, then the factor $f_\ell=1$.  If $\vec{\mathsf{I}}$
is measured in energy units, then $f_\ell =\nu/\nu_\ell^*$. 
Although the radiation transfer equation we use will
retain polarization in general, we will ultimately not include
polarization in computation of Raman source contributions.

There is also an extinction contribution from Raman scattering, which
is given by
\begin{equation} k_{\text{ram}} = \sum_{\ell=0}^{n_R} k_{ram,\ell}=
 \sum_{\ell=0}^{n_R}
n_{H_2}(z)\sigma_{R\ell}(\nu).
\end{equation}
This is the sum over $n_R$ transitions of extinctions involving
photons of incident wavenumber $\nu$ that experience Raman scattering
and emerge as photons of wavenumber $\nu -\Delta\nu_\ell$.  The actual
energy loss at wavenumber $\nu$ due to extinction is proportional to
the incident energy at wavenumber $\nu$, but the source contribution
at wavenumber $\nu$ depends on the incident energy at other higher
wavenumbers.

\subsection{Solution Method}\label{Sec:solmeth}

To solve the vector transfer equation we make use of Fortran vector
radiation code developed and documented by Evans and Stephens (1991)
and further validated by Sromovsky (2004) by comparisons with
independent solutions of Sweigart (1970), Dlugach and Yanovitskij
(1974), and Kattawar and Adams (1971).  Using an order-$m$ Fourier
expansion of the intensity and phase matrix, Eq.\ \ref{Eq:RTEZ} is
separated into $2m+1$ uncoupled equations that can be solved
independently. The azimuthal components are separately solved and then
combined to obtain intensity fields. This separation is valid for
randomly oriented particles with a plane of symmetry, so that the
16-element phase matrix has only six unique components (Hovenier
1969). We also assume that the incident radiation is symmetric in
$\phi$ and unpolarized.  The zenith angle variation is discretized
using a double-Gauss numerical quadrature.  The radiance at any
location in the atmosphere is thus represented as a vector involving
three components: Stokes parameters, quadrature zenith angles, and
azimuthal expansion mode. The intensity and source radiance vectors
thus become vectors of $4N_\mu$ elements, where $N_\mu$ is the
number of zenith angle quadrature points:\begin{equation}
\mathsf{I}= \left[ \begin{array}{c} \vec{\mathsf{I}}(\mu_1)\\ 
\vec{\mathsf{I}}(\mu_2) \\ \cdot
\\ \cdot 
\\ \vec{\mathsf{I}}(\mu_{N_\mu})\end{array}
 \right],
\qquad
\qquad
\mathsf{J}= \left[ \begin{array}{c}
 \vec{\mathsf{J}}(\mu_1)\\ \vec{\mathsf{J}}(\mu_2) \\ \cdot
\\ \cdot 
\\ \vec{\mathsf{J}}(\mu_{N_\mu})
\end{array}\right],
\label{Eq:intvector}\end{equation}
where each element $\vec{\mathsf{I}}(\mu_i)$ or
$\vec{\mathsf{J}}(\mu_i)$ is a vector of the form given in Eq.\
\ref{Eq:stokesvec}. The radiance field is separated into upward and downward
hemispheres: $\mathsf{I^+}$ represents downward radiance ($\mu>0$)
and $\mathsf{I^-}$ representing upward radiance ($\mu<0$).

A model atmosphere is constructed by dividing the atmosphere into
uniform sublayers, each of which is characterized by an optical depth,
a single scattering albedo, and scattering phase function.  The code
uses differential generators derived from Eq. \ref{Eq:RTEZ} and makes
use of doubling to compute reflection and transmission matrices for
homogeneous layers, which are then combined using standard adding
equations.  The key
equations are those of the interaction principle (Goody and Yung
1989): \begin{equation} \begin{array}{c}
\mathsf{I_b^+} = \mathbf{T^+}\mathsf{I_t^+} + \mathbf{R^+}\mathsf{I_b^-} + \mathsf{S^+}\\
\mathsf{I_t^-} = \mathbf{T^-}\mathsf{I_b^-} + \mathbf{R^-}\mathsf{I_t^+} + \mathsf{S^-}
\end{array}\label{Eq:interact}
\end{equation}
where subscripts refer to the top ($t$) and bottom ($b$) boundaries of a
layer, $\mathbf{R}$ and $\mathbf{T}$ are the reflection and
transmission matrices, and $\mathsf{S}$ is the source vector for a
layer, which equals $\mathsf{J}dz/\mu$ for a differential
layer.  Using Eq.\
\ref{Eq:interact}, it is straightforward to derive the properties of a
combined layer ($T+B$) from the individual properties of the top ($T$)
 and bottom ($B$) layers (Evans and Stephens 1991): \begin{equation}
\begin{array}{l l}
\mathbf{R_{T+B}^+ = R_B^+ + T_B^+ \Gamma^+ R_T^+T_B^-}\\
\mathbf{R_{T+B}^- = R_T^- + T_T^- \Gamma^- R_B^-T_T^+}\\
\mathbf{T_{T+B}^+ = T_B^+\Gamma^+ T_T^+}\\
\mathbf{T_{T+B}^- = T_T^-\Gamma^- T_B^-}\\
\mathsf{S_{T+B}^+ = S_B^+} +\mathbf{T_B^+ \Gamma^+}(\mathsf{S_T^+}
                           +\mathbf{R_T^+}\mathsf{S_B^-})\\
\mathsf{S_{T+B}^- = S_T^-} +\mathbf{T_T^- \Gamma^-}(\mathsf{S_B^-}
                           +\mathbf{R_B^-}\mathsf{S_T^+})\\
\mathbf{\Gamma^+ =[1-R_T^+R_B^-]^{-1}}\\
\mathbf{\Gamma^- =[1-R_B^-R_T^+]^{-1}}\\
\end{array}\label{Eq:adding}
\end{equation} 
Here the source adding equations refer to vector sources. 
The same equations can be shown to apply to the matrix
sources that will be defined in a subsequent section. These can be
converted to doubling equations by making the top and bottom
properties equal.

The first step in the solution is to convert Eq.\ \ref{Eq:RTEZ} into a
form that allows computation of the reflection and transmission matrices and
the source vector.  
After azimuthal expansion
and zenith angle discretization, the vector transfer equation for azimuthal
order $m$ can be written in the form\begin{equation}
\begin{array}{c c}
\mathsf{I}_m^{+}(z+\Delta z)= \mathbf{T}_m^{+}\mathsf{I}_m^{+}(z)+
  \mathbf{R}_m^{+}\mathsf{I}_m^{-}(z+\Delta z) + \mathsf{S}_m^{+}\\
\mathsf{I}_m^{-}(z)= \mathbf{T}_m^{-}\mathsf{I}_m^{-}(z+\Delta z)+
  \mathbf{R}_m^{-}\mathsf{I}_m^{+}(z) + \mathsf{S}_m^{-}
\end{array}, \label{Eq:azeq}
\end{equation} where the matrices and source vector here refer to a
 differential layer of thickness $\Delta z$.  The single-scattering
 approximation for the generator equations for reflection and
 transmission matrices are then\begin{eqnarray}
&&|\mathbf{T}^{\pm}_m|_{iji'j'} 
 = \delta_{ii'}\delta_{jj'}\big(1-\frac{k_{ext}\Delta z}
{\mu_j}\big) + \nonumber \\
&&k_{scatt}\Delta z \frac{1+\delta_{0m}}{4\mu_j} w_{j'}
|\mathbf{M}_{m}(z,\nu,\pm \mu_j,\pm\mu_{j'})|_{ii'} \nonumber \\
\nonumber \\
&&\D |\mathbf{R}^{\pm}_m|_{iji'j'} = \nonumber \\
&&k_{scatt}\Delta z\frac{1+\delta_{0m}}{4\mu_j} w_{j'}
|\mathbf{M}_{m}(z,\nu,\pm \mu_{j}, \mp \mu_{j'}) |_{ii'}
\label{Eq:diffgen}
\end{eqnarray}
where $i$ refers to Stokes vector index, $j$ to zenith angle
quadrature index, $w_{j'}$ to the Gaussian quadrature weight value,
and where $\Delta z$ is is usually chosen so that $\Delta \tau =
k_{ext} \Delta z = 10^{-5}$, which Evans and Stephens (1991) have
found to yield about 5 digits accuracy when using double precision.
These expressions are different in form but equivalent to those given
by Evans and Stephens.

The differential generator for Raman source terms can be derived from
the single-scattering form of Eq.\ \ref{Eq:RTEZ} and discretization of
Eq.\ \ref{Eq:source}, which results in the
expression\begin{equation}
\D
\mathsf{S_m^\pm} =
   \sum_{\ell =0}^{n_R}\big[ \mathbf{S}_{\ell,m}^{\pm +}\mathsf{I}_{m}^+(\nu^*_\ell)
  +  \mathbf{S}_{\ell,m}^{\pm -}\mathsf{I}_{m}^-(\nu^*_\ell)\big]
\label{Eq:diffram},\end{equation}
in which the individual matrix elements are given by \begin{eqnarray}
& \D
|\mathbf{S}_{\ell,m}^{\pm \pm}|_{iji'j'}  = & \qquad \nonumber  \\ 
&    f_\ell \times \frac{1}{2} k_{ram,\ell}(z,\nu^*_\ell)\Delta z\frac{w_{j'}}{\mu_j}
    |\mathbf{M}_{R,m}(\pm\mu_j,\pm\mu_{j'})|_{ii'} & 
\label{Eq:sourcemat},\end{eqnarray}
where the first $\pm$ superscript indicates the sign of $\mu_j$ and
the second indicates the sign of $\mu_j'$, and where the Raman
scattering coefficient per unit distance for transition $\ell$ is
given by \begin{equation} k_{ram,\ell}(z,\nu^*_\ell) =
n_{H_2}(z)\sigma_{R\ell}(\nu^*_\ell).\end{equation} Note that the
source expressions depend on the wavenumbers of the incident photons
($\nu_\ell^*$) that are shifted to the current wavenumber, while
everything else in Eq.\ \ref{Eq:azeq} depends on the current
wavenumber $\nu$.  This is not a convenient form to work with because
once the source vector is computed it applies to one specific incident
irradiance direction, meaning that a center-to-limb scan would require
many repetitions of the solution algorithm.  It is also unnecessarily
complex because it retains the full generality of the phase function,
which is not needed to accurately approximate the effect of Raman
scattering.  A simpler and more useful matrix formulation is developed
in Section\
\ref{Sec:smatrix}.

The solution is expressed in terms of reflection and source
matrices for the entire atmosphere, from which intensities at desired
irradiance and view angles are computed as described in Section\
\ref{Sec:compint}. A uniformly spaced wavenumber grid is
used so that the number of photons that are Raman scattered out of one
bin will be transferred to exactly one other bin. This requires that
the wavenumber shift for each transition to be close to an integral
multiple of the grid spacing, as illustrated in Table\ 4. 
Phase functions are introduced as Lagrange function expansions of an
order that depends on the number of quadrature angles.  For
sharply peaked forward scattering phase functions we avoid impractically
large expansion orders, otherwise needed to avoid angular
oscillations in the reflected intensity, by employing the $\delta$-Fit method
of Hu \etal (2000). 

\begin{table*}\centering
\caption{Wavenumber grid spacings that are commensurate with wavenumber
shifts of primary Raman transitions.}
\begin{tabular}{r c c c c c}
   Transitions:  &     $S(0)$   &    $S(1)$  &   $Q+S_1$\\
$\Delta \nu$: &  354.39 cm$^{-1}$ & 587.07 cm$^{-1}$ & 4161.00 cm$^{-1}$\\
\hline
\rule{0pt}{0.15in}
  grid spacing $\delta \nu$ & \multicolumn{3}{c}{$\Delta \nu/\delta \nu$
  (integral steps)}&  RMS error &step sum\\[0.5ex]
\hline
\rule{0pt}{0.15in}
  12.240 cm$^{-1}$ &  28.953 (29) &  47.963 (48) & 339.951 (340) &   0.044 steps& 417\\
  13.640 cm$^{-1}$ &  25.982 (26)  & 43.040 (43) & 305.059 (305) &  0.042 steps& 374\\
  17.780 cm$^{-1}$ &  19.932 (20)  & 33.019 (33) & 234.027 (234) &   0.044 steps & 287\\
  39.260 cm$^{-1}$ &   9.027 (9)  & 14.953 (15) &  105.986 (106) &  0.032 steps & 130\\
  58.620 cm$^{-1}$ &   6.046 (6)  & 10.015 (10) &  70.983 (71)  &  0.029 steps & 87\\
 118.860 cm$^{-1}$ &   2.982 (3)  & 4.939 (5) &   35.008 (35) &    0.037 steps& 43\\[0.5ex]
\hline\label{Tbl:wngrid}
\end{tabular}
\parbox{5in}{NOTE: The step sum is proportional to the number
of storage locations required to save information about shifted
photons when calculating in sequence from shortest to longest wavelengths.}
\end{table*}

While the full polarization machinery of the Evans and Stephens (1991) code
is retained in our modification, the Raman scattering, as we
represent it, does not itself introduce any polarization. 
Further, because in most cases Rayleigh scattering is the dominant
process in creating polarization effects, we usually make use of an
approximation that attributes all of the polarized intensity,
given by $(Q^2+U^2+V^2)^\frac{1}{2}$, to the Rayleigh scattering,
and then use the algorithm described by Sromovsky (2004) to
correct the scalar calculation for polarization, rather than
carry out the full vector calculation.  The approximation
has been shown to be accurate to generally better than 1\%, and
entails no significant additional computational burden, which otherwise
would take about 40 times longer.

\subsection{Matrix Formulation of the Source Function}\label{Sec:smatrix}

A matrix formulation of the source function is used to facilitate
the computation of center-to-limb variations in reflectivity.  Making
the reasonable approximation that Raman scattering is isotropic (see
Sec.\ \ref{Sec:ramphase}), only the $m=0$ azimuthal component
needs to include a Raman contribution and we can write
\begin{eqnarray}
 & |\mathbf{S}_\ell|_{iji'j'} = |\mathbf{S}_\ell^{\pm \pm}|_{iji'j'} = & \nonumber \\
& f_\ell \times \frac{1}{2} k_{ram,\ell}(z,\nu^*_\ell)\Delta
z\frac{w_{j'}}{\mu_j} \delta_{i1}\delta_{i'1} &
\label{Eq:sl} \end{eqnarray}
and simplify Eq.\ \ref{Eq:diffram}  to the form \begin{eqnarray}
& \mathsf{S}^\pm = \sum_{\ell =0}^{n_R} \mathbf{S}_\ell (\mathsf{I^+}(z,\nu^*_\ell)
+ \mathsf{I^-}(z,\nu^*_\ell)) = & \nonumber \\
& \sum_{\ell =0}^{n_R} \mathbf{S}_\ell \mathbf{G}_\ell \mathsf{I^+}(z=0,\nu^*_\ell)
\label{Eq:simplesrc}, &\end{eqnarray} 
where the matrix $\mathbf{G}_\ell$ produces the sum of the upward and
downward intensities at $z$ by multiplication by the incident
intensity at the top of the atmosphere ($z=0$) at wavenumber
$\nu_\ell^*$.  There is an $\ell$ subscript on this matrix because it
depends on transmission, reflection, and source matrices at the
shifted Raman source wavenumber $\nu_\ell^*$.  Expressing the source
in terms of the top-of-atmosphere incident irradiance vector allows us
the computational convenience of expressing the source term as a
matrix multiplier that has the same structure as the reflection
matrix, which will be demonstrated in what follows.

The $\mathbf{G_\ell}$ matrix can be derived from the interaction
principle (Eq. \ref{Eq:interact}).  The radiance at a given level in
the atmosphere is proportional to the incident irradiance and can be
expressed in terms of the properties of the entire atmosphere above
that level (the top layer) and the properties
below that level (the bottom layer). We assume here that the
transmission out of the bottom layer is zero, i.e. if the atmosphere
is thin, the bottom layer includes any surface reflection and/or
absorption. The relevant source vector for the top layer
$\mathsf{S^-_T}$ has multiple terms contributing from different Raman
transitions that are each linearly
related to the irradiance at any single wavelength (the source
direction is the same at all wavelengths).  This can be shown by using
the source addition equations given in Eq.\ \ref{Eq:adding}.  If the
sources for each layer to be combined are separately proportional to
the top of atmosphere irradiance $\mathsf{I^+_0}$
($=\mathsf{I^+}(z=0,\nu_\ell^*)$), then the source for the combined
layer will also be proportional to $\mathsf{I^+_0}$. Thus we can
define source matrices $\mathbf{S^+_T}$ and $\mathbf{S^-_B}$ that
satisfy
\begin{equation}
   \mathsf{S^+_T}(\nu_\ell^*) = 
          \mathbf{S^+_T}(\nu_\ell^*)\mathsf{I_0^+}(\nu_\ell^*)\qquad
   \mathsf{S^-_B}(\nu_\ell^*) = 
          \mathbf{S^-_B}(\nu_\ell^*)\mathsf{I_0^+}(\nu_\ell^*).
\label{Eq:srcprop}\end{equation}
Using these definitions, it is easy to show that \begin{equation}
\mathbf{G_\ell} = (1+\mathbf{R_B^-}) (1-\mathbf{R_T^+
R_B^-})^{-1}\big[ \mathbf{T_T^++S_T^+R_T^+S_B^-}\big] +
\mathbf{S_B^-}\label{Eq:gdef},\end{equation} which defines the matrix
$\mathbf{G_\ell}$ at level $z$ in terms of reflection, transmission,
and source matrices for the entire atmosphere above $z$ (subscript T
for top layer) and that below $z$ (subscript B for
bottom layer), all at wavenumber $\nu_\ell^*$.  The source vector at $z$
for the unshifted wavenumber $\nu$ can then be written as \begin{equation}\D
\mathsf{S^\pm}(\nu) = 
\sum_{\ell=0}^{n_R}\mathbf{S_\ell G_\ell}\mathsf{I_0^+}(\nu_\ell^*)
=\sum_{\ell=0}^{n_R}\mathbf{S_\ell G_\ell}\frac{F_\odot(\nu_\ell^*)}{F_\odot(\nu)}
\mathsf{I_0^+}(\nu)=\mathbf{S}\mathsf{I_0^+}(\nu)
\label{svec}\end{equation}
in which the final two forms define a new matrix $\mathbf{S}$
and express the source contribution to the
current wavenumber $\nu$ in terms of the solar irradiance at $\nu$,
using a scaling that is the ratio of solar spectral functions $F_\odot(\nu)$
and $F_\odot(\nu_\ell^*)$, such that
$\mathsf{I_o^+}(\nu_\ell^*)=[F_\odot(\nu_\ell^*)/F_\odot(\nu)]\mathsf{I_0^+}(\nu)$. If
the radiance field is measured in terms of energy per unit wavenumber,
then $F_\odot(\nu)$ is the solar energy spectrum. If the radiance field is
measured in photons per unit wavenumber, then $F_\odot(\nu)$ is the solar
photon spectrum. Substituting the expression given by Eq. \ref{Eq:sl}
yields the computationally useful form \begin{eqnarray}
(\mathbf{S})_{1j1j''} = \Big[\sum_{\ell} \frac{1}{2} k_{ram,\ell}(z,\nu_\ell^*)
\Delta z \frac{F_\odot(\nu_\ell^*)}{\nu_\ell^*} \nonumber \\
 \sum_{j'} w_{j'} (\mathbf{G_\ell})_{1j'1j''}
\Big] \Big[\frac{\nu}{\mu_j F_\odot(\nu)}\Big]
\label{Eq:srcmult}\end{eqnarray}
in which we have here made a specific choice that $F_\odot(\nu)$ is a
solar spectrum in energy per unit wavenumber, and where $i''=1$ is the
only relevant value of the third index for the assumed unpolarized incident irradiance,
and $i=1$ is the only relevant value of the first index for the assumed isotropic and
Raman scattering phase function, according to Eq.\ \ref{Eq:sl}.  The
$\pm$ superscript on $\mathsf{S}$ was dropped, and never used on
$\mathbf{S}$, because both values are the same for a differential
layer (see Eq.\ \ref{Eq:simplesrc}). Note that the first bracketed
factor depends on $\nu_\ell^*$ while the second bracketed factor
depends only on wavenumber $\nu$ and $z$, and is independent of
$\ell$.  This form suggests the computational procedure described below.

\subsection{Computation of the Source Matrix}

Atmospheric layers that have uniform extinction coefficients and phase
functions become nonuniform due to the vertical variation of the Raman
source contribution.  This requires us to use a much larger number of
layers than would be necessary were Raman scattering not considered.
We typically use 30-80 logarithmically spaced pressure levels between
0.0003 and 100 bars, with 80 levels required for accuracy of a few
tenths of 1\%.  The accuracy achieved in any particular problem needs
to be assessed by running test calculations using a larger number of
levels.

Each Raman solution requires a spectral calculation. Following
 B\'{e}tremieux and Yelle (1999),  we
start at the highest wavenumber and work downward so that multiple
Raman scattering is automatically included in a single pass. At the highest
wavenumber, the radiation transfer problem is solved with no
source contribution, but with Raman extinction included.  At each
wavenumber including the first, the $\mathbf{G}$ matrix is computed at
each layer bottom boundary $L$, using Eq.\
\ref{Eq:gdef}.  The algorithm then computes for each transition $\ell$, each layer
$L$, and each possible incident quadrature index $j''$, the function
\begin{eqnarray} H(L,\ell,j'') = k_{ram,\ell}^L(\nu \rightarrow
\nu-\Delta\nu_\ell)  \frac{F_\odot(\nu)}{\nu} \nonumber \\
\sum_{j'}w_{j'}(\mathbf{G^L+G^{L+1}})_{1j'1j''}
\label{Eq:srcstore}\end{eqnarray}
which is essentially the same as the left bracketed factor in Eq.\
\ref{Eq:srcmult} divided by $\Delta z$. This is essentially the
photon loss term per unit distance for each layer and each transition.
Here everything is evaluated at the same wavelength and thus the  $\mathbf{G}$
matrix needs no $\ell$ dependence. This form uses the average of $\mathbf{G}$ at the
top and bottom of each layer to represent the
radiation field everywhere within the layer.

As the computation proceeds to longer wavelengths, the source
contributions at the current wavenumber are then obtained by reading
the stored values of $H$ and computing the source matrix as
follows:\begin{equation} \mathbf{(S^L)_{1j1j''}} =
\Big[\sum_{\ell} H(L,\ell,j'') \Big]
\frac{\nu}{4\mu_j F_\odot(\nu)} \Delta z
\label{Eq:finalsrc}.\end{equation}
For each
wavenumber at which values of $H$  are read, the values for transition $\ell$
will have been stored during calculations made at wavenumber $\nu + \Delta\nu_\ell$.

Once $H$ is read from storage and the matrix $\mathbf{S^L}$ is
computed, the computation of the radiation field proceeds with the
assumption that the source is uniformly distributed over the layer
$L$.  This is necessary to make use of the doubling equations for a
uniform source, but is only a good approximation for layers that are
relatively optically thin at the wavenumber of the incident photons.
While the input source is assumed to be uniform throughout each layer,
and is input as the differential source after dividing by the physical
thickness of each layer, the doubling process does modify the source
function in accord with the absorption and scattering processes that
take place at the wavelength of the scattered Raman photon. 

\subsection{Computation of Intensity}\label{Sec:compint}

After computing the source matrix for the entire atmosphere using the
same adding equation that is given for the source vector in Eq.\
\ref{Eq:adding}, we then compute the outward intensity at the top of the
atmosphere.   For arbitrary directions of incidence and reflection this
is formally expressed as \begin{eqnarray} \mathsf{I_0^-}(\mu,\phi) &=& 
\Big[\mathbf{R^-}(\mu,\mu_0,\phi-\phi_0) +
\mathbf{S^-}(\mu,\mu_0,\phi-\phi_0)\Big] \nonumber \\
&&\times \left[\begin{array}{c} \mu_0 F_0/\pi \\ 0 \\ 0 \\ 0\end{array}\right]
=\mathbf{\bar{R}^-}\left[\begin{array}{c} \mu_0 F_0/\pi \\ 0 \\ 0 \\ 0\end{array}\right]
\label{Eq:toaint}\end{eqnarray}
where the incident flux is $F_0$ in direction ($\theta_0, \phi_0$) and
the combined matrix $\mathbf{S^-}$ uses the minus superscript because
direction does make a difference for the inhomogeneous layer consisting
of the entire atmosphere.  By converting the
source function into a matrix we have thus put the Raman
contributions in the same form as the reflection contributions,
allowing us to compute a pseudo reflection matrix
$\mathbf{\bar{R}^-}$ and  use the same mathematical machinery to
compute spatially resolved intensity profiles.

The azimuthal expansion of $\mathbf{\bar{R}^-}$ is given by \begin{eqnarray}
\mathbf{\bar{R}^-}(\mu,\mu_0,\phi-\phi_0)Z= 
 \mathbf{\bar{R}^-_0}(\mu,\mu_0)\nonumber &&\\+\sum_{m=1}^{\infty}
\mathbf{\bar{R}_m^-}(\mu,\mu_0)\cos(m(\phi-\phi_0))&&
\label{Eq:azexpofr}\end{eqnarray}
Where sine components of the expansion are zero by the assumed
symmetry of the incident radiation field. Since our solution only
provides $\mathbf{\bar{R}^-}$ at quadrature points, we carry
out a Legendre polynomial interpolation to obtain intensities at other angles.

\subsection{Validation}

How do we know that our computations are correct?  The basic radiation
transfer algorithm without Raman scattering has been compared with
independent calculations for selected cases and shown to be in
excellent agreement.  This was done by Evans and Stevens (1991) for
their original code and by Sromovsky (2004) for the modification used
here (as noted in Sec. \ref{Sec:solmeth}), but neither of these sets
of comparisons deals with the Raman scattering additions, which are
harder to validate.  A few comparisons with past calculations can be
made, but most of these lack the resolution, relevance, or rigor that
are needed to serve as appropriate standards of accuracy.  They do
provide a reasonable sanity check, however, and some useful
comparisons are made in Section\ \ref{Sec:compcal}.  Another
validation is to show by means of test cases that the algorithm
satisfies conservation of photons. This is done for the case of a
monochromatic incident flux in Section\
\ref{Sec:monoscat}. Conservation is unlikely to be achieved if the
Raman transfers are improperly computed.  A last validation, though an
indirect one, is to demonstrate that the algorithm can reproduce
features in the observed spectra with reasonable atmospheric structure
models (Sec.\ \ref{Sec:obscomp}).

\section{Atmospheric and Solar Models}

\subsection{Model Atmosphere of Neptune}

Neptune's thermal structure is obtained from Voyager radio occultation
observations at ingress (45\deg S) and egress at 61\deg N). We used
the Hinson and Magalh\~aes (1993) analysis for $p < 1$ bar and the
Lindal (1992) results for 1 bar $< p< 6$ bars, with an offset of 1.0 K
added to match the the Hinson and Magalh\~aes profile at 1 bar.  For
$p > 6$ bars, we extrapolated using the nearly adiabatic lapse rate at
the 6-bar level ($\approx$ -0.94 K/km).  The two temperature profiles
differ insignificantly except in the stratosphere.  The lower egress
temperatures in the tropopause region reduce the CH$_4$ mixing ratio
and opacity to a small degree that is only noticed near 0.89 $\mu$m,
where a 10\% increase in geometric albedo is produced.  We assume a
fixed profile at all latitudes for radiation transfer analysis.  Our
altitude scale is computed using gravitational acceleration at 45\deg
S ($g=11.1$ m/s$^2$).  The Hinson and Magalh\~aes profile is derived
assuming a gas composition of 81\% H$_2$ and 19\% He (Conrath \etal
1991).  It is possible that N$_2$ may also be present at a mixing
ratio as high as 0.3\% (Conrath \etal 1993), in which case our
altitude scale would be modified somewhat. The Lindal profile assumed
2\% CH$_4$ in the troposphere, which is close to the currently
accepted value of 2.2\% (Baines \etal 1995) that we use for computing
CH$_4$ opacity.  The tropospheric mixing ratio is consistent with
CH$_4$ condensation at 1.4 bars, above which we assume the saturated
vapor pressure until we reach the stratosphere, at which point the
mixing ratio is the smaller of the saturated mixing ratio and the
stratospheric limit of 3.5 $\times 10^{-4}$ (Baines and Hammel 1994).

\begin{figure}[!htb]
\hspace{-0.2in}\includegraphics[width=3.7in]{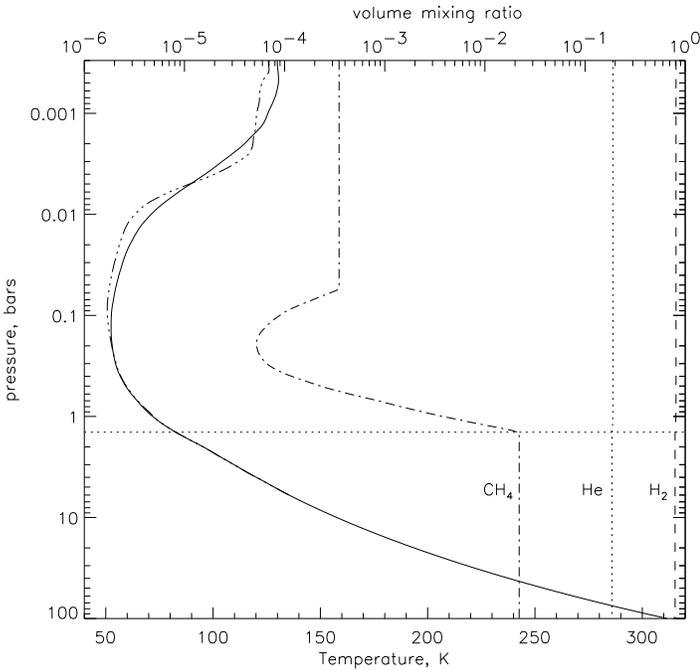}
\caption{Vertical temperature structure of Neptune based on
ingress (solid) and egress (triple dot dash) occultation observations,
and volume mixing ratios of H$_2$ (dashed), He (dotted), and CH$_4$
(dot dash). The CH$_4$ mixing ratio is 3.5 $\times 10^{-4}$ in the
stratosphere, 0.022 in the deep troposphere, and otherwise follows the saturation
vapor pressure curve. The horizontal dotted line at 1.43 bars indicates
the point at which methane reaches the saturation vapor pressure in
the troposphere.}
\label{Fig:ingressmix}
\end{figure}

\begin{figure}[!htb]
\hspace{-0.2in}\includegraphics[width=3.7in]{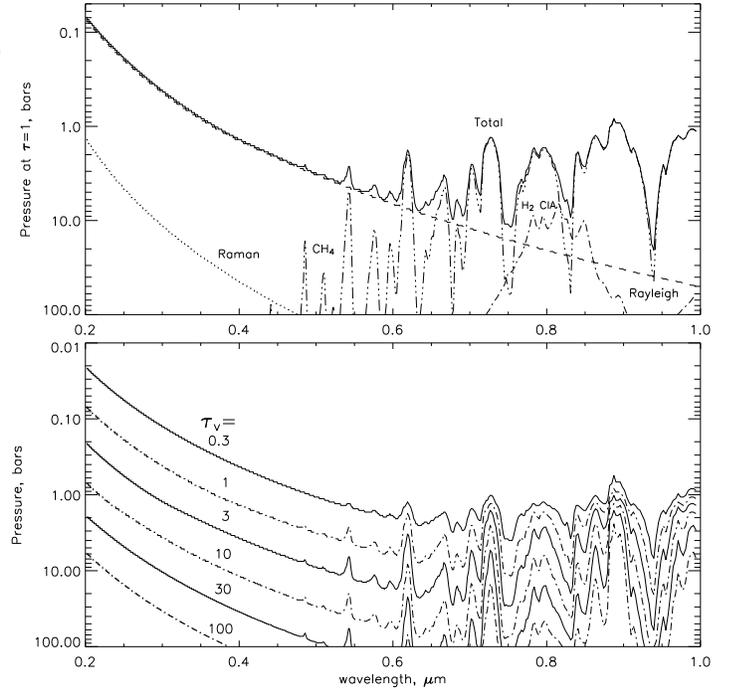}
\caption{Top: Pressure level at which Rayleigh scattering (dashed), 
Raman scattering (dotted), CH$_4$ absorption (triple dot-dash), H$_2$
CIA (dot-dash), and the sum of all opacities (solid) accumulate a
one-way vertical optical depth of unity, plotted versus wavelength.
Bottom: Pressure at which one-way total vertical optical depth reaches
0.3, 1, 3, 10, 30, and 100, plotted versus wavelength.}
\label{Fig:pendepth}
\end{figure}

 The pressure at which the one-way vertical optical depth reaches
 unity for each opacity contributor, and for the total of all
 contributors is illustrated in the top panel of Fig.\
 \ref{Fig:pendepth}. The levels at which the total optical depth
 reaches values from 0.3 to 100 are displayed in the bottom panel. The
 penetration depth of sunlight into Neptune's atmosphere is limited by
 Rayleigh scattering at short wavelengths and by CH$_4$ and H$_2$
 collision-induced absorption (CIA) at long wavelengths. The deepest
 penetration is at 0.935 $\mu$m, where there is a relatively clean
 CH$_4$ window, a window in the CIA spectrum, and a low Rayleigh cross
 section.  We used CH$_4$ absorption coefficients derived by
 Karkoschka (1994) from planetary geometric albedo observations using
 a technique described by Karkoschka and Tomasko (1992). Karkoschka
 estimates a 5\% uncertainty in his 1994 absorption coefficients, plus
 an additional uncertainty in the continuum baseline: 0.0003
 km-am$^{-1}$ at 400 nm and 0.02 km-am$^{-1}$ at 1000 nm (a factor of
 2 every 100 nm). This continuum uncertainty leads to uncertainties in
 the CH$_4$ window regions that are important to determination of the
 scattering properties and pressure levels of clouds in the several
 bar range. There is also a likely bias error for weak CH$_4$ bands
 superimposed on Raman scattering effects due to the nonlinear
 relationship between I/F and single-scattering albedo (this is
 discussed in Sec.\ \ref{Sec:kcorrect}). The CIA values are obtained
 by interpolating tables of pressure and temperature dependencies
 provided by Alexandra Borysow, and available at the Atmospheres Node
 of NASA'S Planetary Data System. The average Rayleigh scattering
 cross section per molecule was computed using the equation
 \begin{equation}
\sigma_\mathrm{Ray} = \frac{8\pi^3}{3\lambda^4N^2}\sum_i
v_i(n_{g,i}-1)^2
\frac{6+3\delta_i}{6-7\delta_i}
\label{Eq:rayxc}\end{equation} 
from Hansen and Travis (1974), where $v_i$ is the volume mixing ratio
of the $i$th gas, $n_{g,i}$ is its refractive index, and $\delta_i$ is
its depolarization factor. We used depolarization ratios of 0.0221 for
H$_2$ and 0.025 for He (Penndorf 1957; Parthasarathy 1951) and
refractive index values from Allen (1964). Because no depolarization
values were give for CH$_4$, we used CO$_2$ values. We also used the
wavelength dependence of ammonia's refractive index to approximate
that of CH$_4$. The error in these latter approximations
is not significant because of the small CH$_4$ mixing ratio.

\subsection{Cloud and Haze Models}\label{Sec:cloudandhaze}

\begin{figure}[!htb]
\includegraphics[width=3.4in]{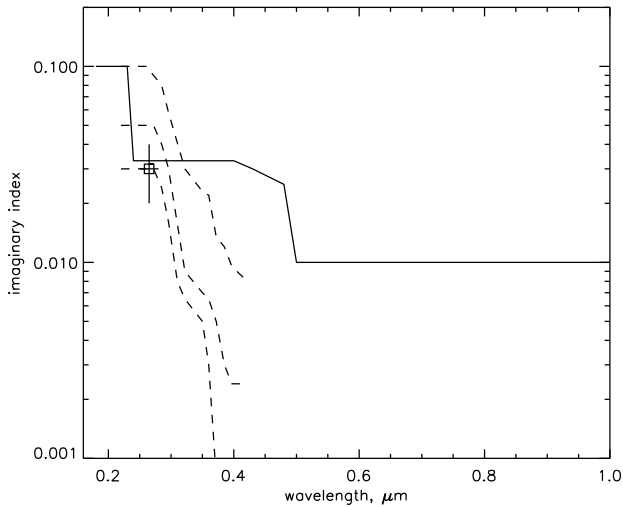}
\caption{Upper haze imaginary index assumed here (solid curve)
compared to that inferred by Courtin (1999) (dashed curves) and by
Pryor \etal (1992) (plotted point).}
\label{Fig:nimag}
\end{figure}

Two model structures with aerosols were chosen to illustrate the
effects of stratospheric haze and lower altitude cloud aerosols on
Raman scattering features and to test approximations under more
realistic conditions than provided by a clear atmosphere. Both models
contain a high-altitude absorbing haze of Mie-scattering spherical
particles with a Hansen (1971) gamma size distribution of effective
radius $a$ = 0.2 $\mu$m and variance $b$ = 0.02, where the relative
number per unit radius interval at radius $r$ is proportional to
$r^{(1-3b)/b}\exp(-r/(ab))$. The particle size is from Pryor \etal
(1992). The refractive index of the haze is assumed to have a real
value of 1.44 and $\lambda$-dependent imaginary values given in Fig.\
\ref{Fig:nimag}, which differ from the variation
inferred by Courtin (1999).  Other index variations could also have
been used, with compensating changes in other aerosol
components. Finding a tightly constrained fit is left for future work.
While Courtin's $\tau$ = 0.1 haze extends downward only to 20 mb, our
haze has twice the optical depth and extends much more deeply; it has
a uniform mixing ratio between 100 and 800 mb, which is more similar
to the distribution inferred by Moses
\etal (1995). Both of our haze
models also contain a 3.8-bar cloud of isotropic scatterers, with a
single-scattering albedo 0.99.  This cloud is at the level that is
often considered to be the top of a semi-infinite cloud (Baines \etal
1995).  Haze Model I has this deep cloud set to optical depth 0.5.
For Haze Model II this cloud is set to unit optical depth and a second
cloud is placed at 1.3 bars, the approximate level expected for the
base of a CH$_4$ ice cloud assuming a CH$_4$ mixing ratio of
2.2\%. For this cloud we assumed $\tau$ = 2, a single-scattering
albedo of 0.99, and a double Henyey-Greenstein phase function of the
form $P(\theta)$ = $f_1 P_1(\theta) + (1-f_1) P_2(\theta)$, with
$g_1=0.9$, $g_2=-0.11$, and $f_1=0.42$, where
$P_i(\theta)=(1-g_i^2)/(1+g_i^2-2g_i
\cos(\theta))^{3/2}$. The phase function parameters are due to Pryor
\etal (1992), based on high-phase angle Voyager images. Our haze and
cloud structures are not meant to be optimum fits to Neptune's
geometric albedo spectrum, but rather a sampling of possible structures
that might be encountered.  Nevertheless, the part of the spectrum
below 0.5 $\mu$m is a fairly good match to the observed geometric
albedo spectrum, as demonstrated in Sec. \ref{Sec:obscomp}. The
Haze II model mainly serves as a test case for approximations.

\subsection{Models of the Solar Irradiance Spectrum}

\begin{figure*}[!htb]\centering
\includegraphics[width=6.4in]{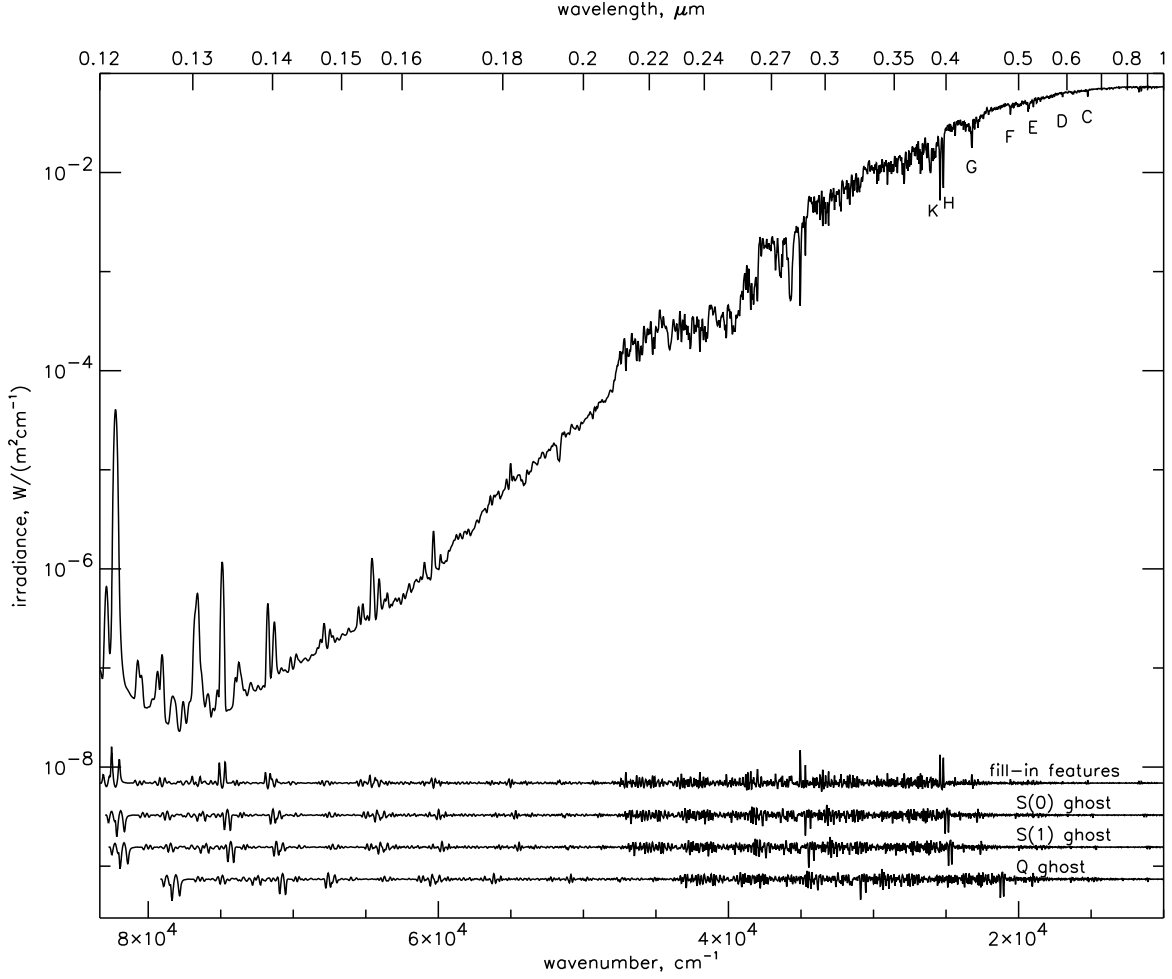}
\caption{Solar irradiance spectrum in energy flux units per wavenumber,
using the UARS spectrum (Woods \etal 1996) for wavelengths less than
0.41 $\mu$m, and a normalized model spectrum for longer wavelengths
(Kurucz, 1993b), convolved to a wavenumber resolution of 35 cm$^{-1}$
and sampled at twice that frequency. The letters C-K label the more
prominent Fraunhofer absorption lines. At the bottom are scaled and
offset ratio plots illustrating fill-in features (smoothed
spectrum/unsmoothed), $S(0)$ ghosts (unsmoothed/smoothed with 354.4
cm$^{-1}$ wavenumber shift), $S(1)$ and $Q$ ghosts (unsmoothed /smoothed
with 587cm$^{-1}$ and 4161 cm$^{-1}$ shifts respectively).}
\label{Fig:solarspec}
\end{figure*}

A key input to modeling of Raman scattering is an accurate solar
spectrum of the appropriate spectral resolution. 
For $\lambda > 0.41$ $\mu$m we used a model spectrum (Kurucz 1993)
normalized to the 0.41 to 0.87 $\mu$m results of Neckel and Labs
(1984) and for 0.12 $\mu$m $< \lambda <$ 0.41 $\mu$m we used
measurements by the Upper Atmospheric Research Satellite using the
SOLSTICE instrument (Woods \etal 1993).  The UARS spectrum was
obtained from the UARS web site
(ftp://rescha.colorado.edu/pub/solstice/sol\_hires\_200.dat). It has a
nominal resolution of 0.2 nm, but is actually closer to a resolution
of 0.5 nm, based on comparisons with convolutions of the very high
resolution Kurucz model spectrum.  
We created our standard solar
reference spectrum (Fig.\ \ref{Fig:solarspec}) by convolving the
combined spectrum with a triangular sampling function to obtain a
nominal FWHM resolution of 0.35 cm$^{-1}$ sampled at 0.1725
cm$^{-1}$.  The nominal wavelength resolution varies from 0.14 nm at 0.2
$\mu$m to 3.5 nm at 1 $\mu$m (the resolution is 1 nm at 534 nm).
Although this has nominally a uniform wavenumber resolution, it is
actually limited by the UARS observations to about 0.5 nm at the
shortest wavelength, where it is oversampled by a factor of 2 or more.
The spectrum doesn't actually reach the nominal resolution until the
transition to the Kurucz spectrum at 0.41 $\mu$m.  The solar spectrum
matches the Karkoschka ground based observing resolution of 1 nm at
0.534 $\mu$m, but is 3.5 times worse near 1 $\mu$m. Thus careful
comparisons between observation and model calculations need to adapt
to differing resolutions.

Following Courtin (1999), we selected a UARS spectrum from 29 March
1992 to provide optimum compatibility with the FOS observations from
19 August 1992.  Courtin found a peak-to-peak difference of only 2\%
between two low-resolution UARS daily spectra obtained on 2 June 1992
and 19 August 1992. Thus, solar variability is unlikely to be a major
error source in comparing calculations with FOS observations. Most of
the solar variability is restricted to $\lambda <$ 0.26 $\mu$m, so
that comparison with the groundbased observations made in 1993
(Karkoschka, 1994) are even less influenced by solar variability.

\section{Characteristics of Raman Spectra}

\subsection{Raman Scattering for a Monochromatic Source}\label{Sec:monoscat}

\begin{figure*}[!htb]\centering
\includegraphics[width=6.in]{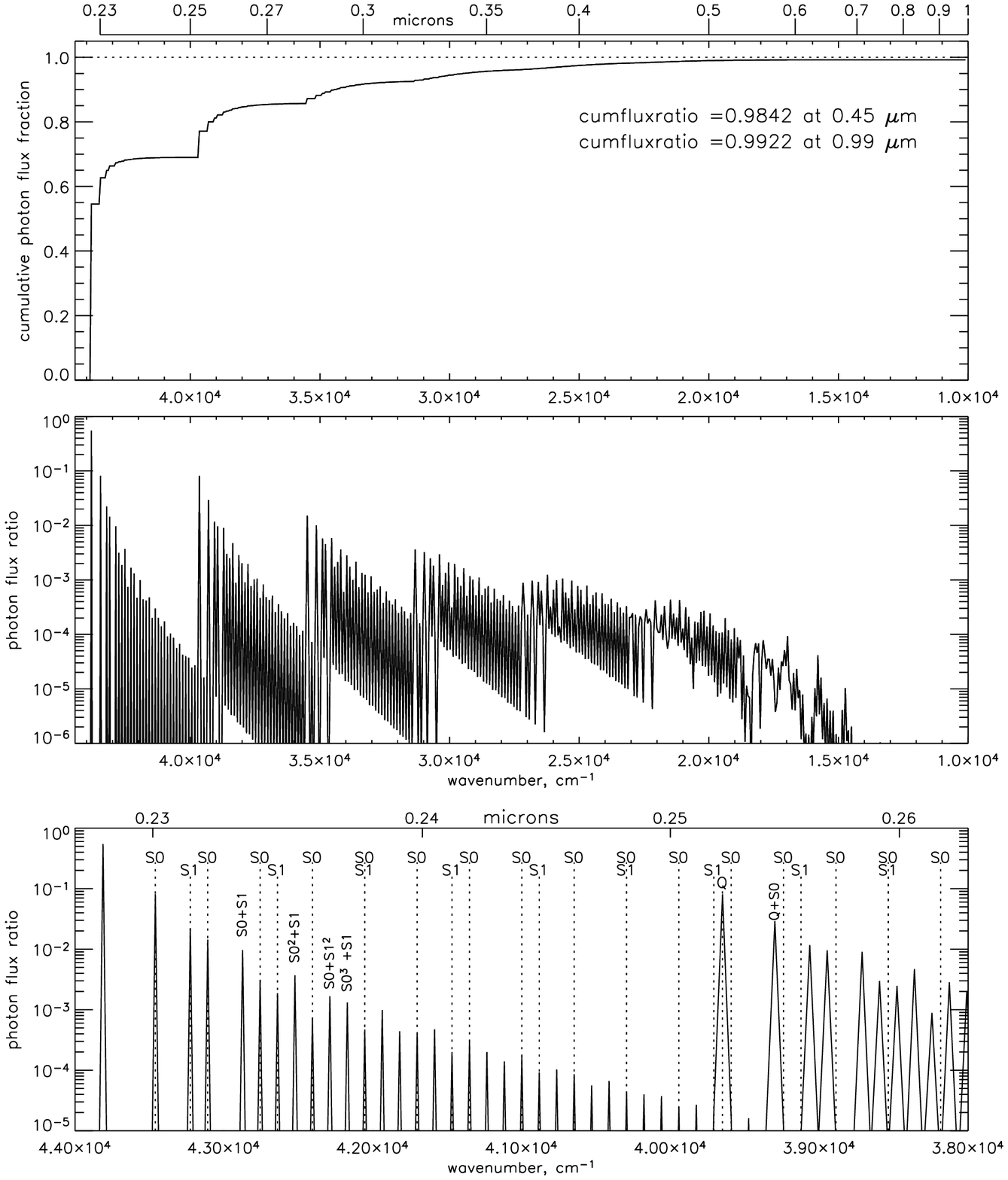}
\caption{Bottom: Photon flux ratio vs wavenumber for an incident monochromatic flux
at 43812.2 cm$^{-1}$ (228.247 nm). Selected flux peaks are labeled by
Raman transition or combination of transitions that produced them. Middle: extended
version of bottom panel. Top: Cumulative fractional photon flux.}
\label{Fig:ramline1plus2}
\end{figure*}

The behavior of monochromatic incident photons provides useful insight
into the workings of Raman scattering in Neptune's atmosphere and a
useful validation of our basic computational algorithm.  For this
example we use the three Raman transitions given in Table\ 4, a
wavenumber spacing of 58.62 cm$^{-1}$, an atmosphere with CH$_4$
absorption and H$_2$ CIA, an equilibrium distribution
of H$_2$, but no aerosols.  
The propagation of photons
from the injection wavelength of 0.228247 $\mu$m (43812.2 cm$^{-1}$)
is illustrated in the bottom panel of Fig.\
\ref{Fig:ramline1plus2}.  The vertical scale in this plot is the ratio of
the total reflected flux at normal incidence to the incident photon
flux. The first peak shows that 54.6\% of
the incident photons are reflected by the atmosphere at the incident
wavelength. The remaining 45.4\% of the photons are shifted to longer
wavelengths by Raman scattering.  The second peak, containing about
8.1\% of the photons is due to the $S(0)$ transition. This second peak
is the fraction of single-scattered $S(0)$ photons that exit the
atmosphere; but many $S(0)$ photons are further shifted to even longer
wavelengths.  The third peak, containing 2.2\% of the incident
photons, is due to the $S(1)$ transition. Many additional peaks are
due to multiple Raman scattering with various combinations of $S(1)$
and $S(0)$.

The relatively large peak near 0.252 $\mu$m, containing 8.1\% of the
incident photons, is the first due to the $Q$ transition. This is the
same size as the $S(0)$ peak and thus seems surprisingly large given
that the $Q$ transition has about half cross section of the $S(0)$
transition (the ratio is 0.541 at 0.228 $\mu$m).  However, the $S(0)$
contribution is multiplied by $f_{para}$, which is below 0.75 in the
pressure range likely to contribute most at this wavelength (Fig.\
\ref{Fig:parapop}).  One might thus expect the $Q$ contribution to be
at least 72\%, of the $S(0)$ contribution on this basis
alone. However, there are other factors also at work, such as multiple
scattering and the vertical distribution of Raman photons. In fact, if
the Rayleigh scattering cross section is made wavelength-independent,
then the $Q$ peak becomes only about 57\% of the $S(0)$ peak, which is
less than expected from cross section and $f_{para}$ considerations.
Apparently, in the real atmosphere the decreasing opacity of the
atmosphere with increasing wavelength allows more of the Raman
photons to make it out of the atmosphere than would otherwise be the
case. Because the $Q$ photons undergo the largest wavelength shift,
they exhibit the largest effect.

The conservation of photons illustrated in the top panel of
Fig.\ \ref{Fig:ramline1plus2} is computed as follows.
For an incident unpolarized flux $F_0(\nu)$ at zenith cosine $\mu_N$, the
quadrature value closest to unity, the total reflected flux at $\nu$ is given by
by the discrete summation
 \begin{eqnarray}
& F_r (\nu) = 
\Big[2\sum_{k=1}^{N} w_k \mu_k \Big( \mathbf{R^-(\nu)}_{k1N1} \nonumber & \\
&+ \mathbf{S^-(\nu)}_{k1N1}\Big)\Big]\mu_N F_0(\nu), &
\label{Eq:reftot}\end{eqnarray}
where, for a monochromatic incident flux at $\nu=\nu_0$, only the
reflection term contributes at $\nu=\nu_0$ and only the source term
contributes at $\nu<\nu_0$.  Because of the matrix formulation of the
source term that is used here, we must introduce a
pseudo-monochromatic flux that has a negligibly small constant offset
for $\nu\ne\nu_0$, to avoid dividing by zero in Eq.\
\ref{Eq:finalsrc}. Using discretization of wavelength as well, in
which the incident wavenumber is $\nu_0$ and $\nu_j = \nu_0
-j\Delta\nu$, we can write the pseudo-monochromatic incident flux as
$F_0(\nu_j) \times (\delta_{0,j} + \epsilon)$ implying a photon line
flux (flux at wavenumber $\nu_0$) of $F_0(\nu_0)/(h\nu_0)$.  The
cumulative fractional photon flux can then be written as
\begin{eqnarray} 
&f_{cum}(J) =\frac{\nu_0}{F_0(\nu_0)}\sum{j=0}^{J}
 \frac{F_r(\nu_j)}{\nu_j}
 = & \nonumber \\
& 2\sum_{k=1}^{N} w_k \mu_k  \mathbf{R^-(\nu_0)}_{k1N1}\nonumber & \\
&+ 2\epsilon\sum_{j=1}^{J}\sum_{k=1}^{N} w_k \mu_k
 \mathbf{S^-(\nu_j)}_{k1N1}\mu_N
\frac{\nu_0 F_0(\nu_j)}{\nu_j F_0(\nu_0)},&
\label{Eq:sumflux} \end{eqnarray}
where $J < \nu_0/\Delta\nu$ and where
we ignored $\epsilon$ compared to unity in the first term and
made use of the fact that the source term is zero at $\nu=\nu_0$.  Although
the second term has an explicit multiplication by $\epsilon$, it is
not negligibly small because the source term itself contains a
division by $\epsilon$. For a conservative atmosphere that is either
semi-infinite or bounded by a unit-albedo surface, $f_{cum}$
 should evaluate to unity if the sum is extended to the last wavenumber.
The cumulative flux shown in Fig.\ \ref{Fig:ramline1plus2} falls somewhat
short of unity because of CH$4$ absorption and H$_2$ CIA.

The middle panel of Fig.\ \ref{Fig:ramline1plus2} shows that Raman
photons can undergo a surprisingly large number of scatterings before
they leak out at the top of the atmosphere. In fact, as shown by the
plot of cumulative flux in the upper panel, even by 0.45
$\mu$m, 1.6\% of the incident photons are still inside the
atmosphere. This occurs because the atmosphere is conservative at short
wavelengths and
because photons scattered deep within the atmosphere are more easily
lost to another Raman scattering than lost to transmission out of the
atmosphere (to space).  After each additional Raman scattering, the
optical depth to space gets smaller, as does the cross section for
further Raman scattering (Fig.\ \ref{Fig:xcvswlen}), so that photons
are more easily lost to space at longer wavelengths.

\begin{figure}[!htb]
\hspace{-0.15in}\includegraphics[width=3.7in]{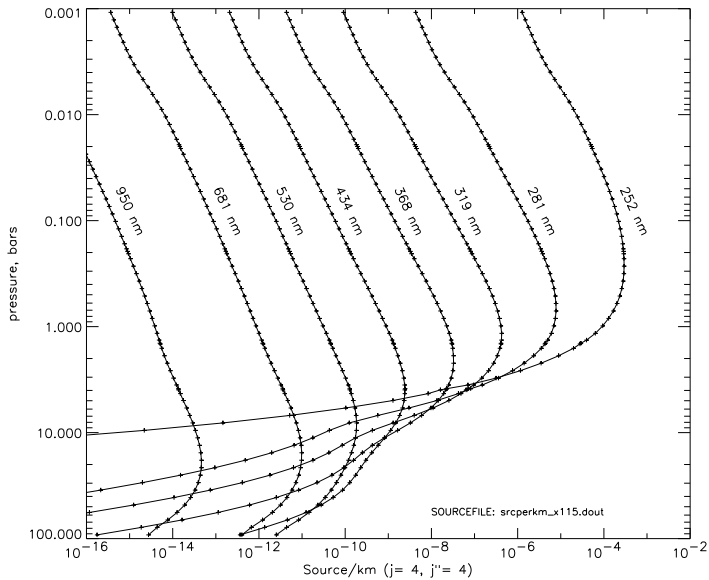}
\caption{Relative source function intensity $S_L(j,j'')$ per km versus 
geometric mean pressure of the layer boundaries, for selected
wavelengths (labeled by nanometers), for a clear Neptune atmosphere.}
\label{Fig:linesrcperkm}
\end{figure}

The vertical distribution of Raman scattered photons is illustrated in
Fig.\ \ref{Fig:linesrcperkm} for the $Q$ transition, which produces
peaks spaced by 4161 cm$^{-1}$ (Fig.\ \ref{Fig:ramline1plus2}).  The
curve of highest magnitude is due to one scattering, the second due to
two scatterings, and the last is due to eight successive scatterings
involving the $Q$ transition.  Because of the large wavenumber shift
of this transition, these peaks decline more slowly than those related
to the $S(0)$ and $S(1)$ transitions.  The peak at the shortest
wavelength is at a location where the incident radiation field times
atmospheric density is at a local maximum, which occurs near 300 mb.
After each additional scattering, the Raman photons diffuse both
upward, where they leak out of the atmosphere, and downward, where
they provide a reservoir for later contributions.
It is worth noting that some of the Raman scattered photons that
are transferred to a longer wavelength can leave
the atmosphere directly without any further scattering (those at
high altitude and directed upward).  The result is
a movement of the peak source intensity further downward after each
successive scattering.

Because of vertical variation in the source function, it was necessary
to use 72 logarithmically spaced gas layers to avoid cumulative photon
fluxes exceeding unity. When cross sections were made wavelength
independent, but evaluated at 0.4 $\mu$m, the accuracy for a given
number of levels was much higher. Monochromatic calculations were run
with three different choices for the number of logarithmically spaced
atmospheric levels between 0.0003 and 100 bars: 18, 36, and 72.  The
ratios to the 72-level fluxes, were 0.991-0.998 for the 18-level
result, and 0.998-0.9996 for the 36-level result.  Thus it seems
likely that errors for 72 levels probably do not exceed 0.1\%, and 36
levels are probably within 0.2\%.  The ratios to the 72-level
cumulative flux at 1 $\mu$m were 0.9976 for 18-levels and 0.9995 for
36 levels.

\subsection{Raman Scattering of a Flat Spectrum with an Absorption Line}
\begin{figure}[!htb]
\hspace{-0.2in}\includegraphics[width=3.7in]{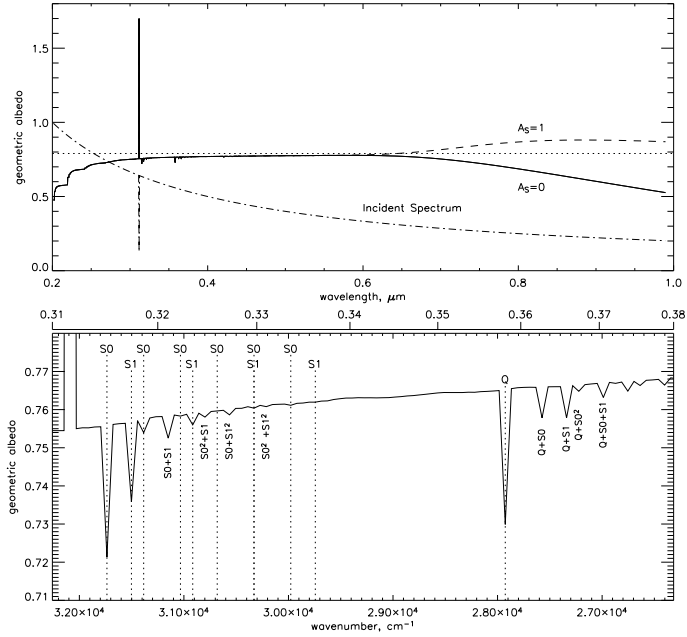}
\caption{Upper: Flat photon irradiance spectrum (dot-dash) with an absorption
feature at 0.311641 $\mu$m (32088 cm$^{-1}$) and corresponding geometric
albedo spectra of Neptune for unit (dashed) and zero (solid) surface
albedos. The dotted line indicates the geometric albedo for a
Raman-free semi-infinite Rayleigh atmosphere (p=0.7908). Lower:
Geometric albedo vs wavenumber, showing Raman transition assignments
for the main ghost features.}
\label{Fig:flatphotabs}
\end{figure}

For an irradiance spectrum with a $\lambda$-independent photon flux, one
might expect the apparent geometric albedo with Raman scattering to be
the same as without it.  One could argue that for every photon lost
from a given wavenumber, there is one gained from a higher
wavenumber. This argument works well if atmospheric properties are
also independent of wavelength. But in a real atmosphere, Raman and
Rayleigh scattering cross sections vary with wavelength, so that the
vertical distribution and fraction of photons lost will be somewhat
different from the vertical distribution and fraction of photons
gained, even with a $\lambda$-independent incident photon flux.  In
spite of this complication, the results shown in Fig.\
\ref{Fig:flatphotabs} are roughly consistent with this argument.
The calculation shown here is for a clear Neptune atmosphere with
CH$_4$ absorption and CIA turned off, and exposed to an irradiance
spectrum that has a constant photon flux except for a pseudo-solar
absorption feature at 0.311641 $\mu$m (32088 cm$^{-1}$), where the
photon flux was dropped to 20\% of its value elsewere. The irradiance
plotted here is the spectrum of energy per unit wavenumber, normalized
to unity at 0.2 $\mu$m, which corresponds to a flat photon spectrum.
Note the considerable distance from the starting wavelength to the wavelength
at which the geometric albedo approaches the Raman-free semi-infinite
H$_2$ Rayleigh value of 0.7908 (Sromovsky 2004). That is not
surprising because of the importance of multiple Raman scattering
evident from Fig.\
\ref{Fig:ramline1plus2}. Calculation were run with surface albedos of 0
and 1, to show that for $\lambda >$ 0.6 $\mu$m even 100 bars of
atmosphere is not enough to obscure the lower boundary when only
Rayleigh and Raman scattering are considered.  With CH$_4$ absorption
turned on, the effects of the bottom boundary are no longer apparent.

Among the sharp spectral features present in the reflection spectrum,
the largest peak is at exactly the wavelength of the absorption spike
in the irradiance spectrum.  This peak arises because in addition to
photons reflected at the same wavelength, there are also photons Raman
shifted from shorter wavelengths.  Because more are shifted in than
shifted out, sharp absorption features like this get partially filled
in the reflected flux.  When the reflected flux is then divided by the
irradiance to obtain the albedo, the feature stands out as a positive
spike. Solar absorption features like this are the most prominent
features in the Raman spectra of the outer planets.  The other, much smaller
downward spikes are ghosts of the irradiance spectrum, displaced by the
wavenumber shift of the Raman transition.  In the detailed view shown in
the lower part of Fig.\ \ref{Fig:flatphotabs} the ghost features are
labeled by the transitions that produced them. Note the relatively
small size of the ghost features.  The $S(0)$ ghost feature has an
amplitude of about 5.8\% of the geometric albedo, while the
corresponding absorption feature in the irradiance spectrum has an
amplitude of 80\%; the $S(1)$ and $Q$ features have amplitudes of
about 2.7\% and 4.8\% respectively. The near equality of $Q$ to $S(0)$
ghost amplitudes in spite of their very different cross sections has
the same explanation as the relatively large $Q$ contribution peak in the
monochromatic spectrum, discussed in the previous section.

\subsection{Raman Scattering of the Solar Spectrum}

\begin{figure}[!htb]
\hspace{-0.2in}\includegraphics[width=3.7in]{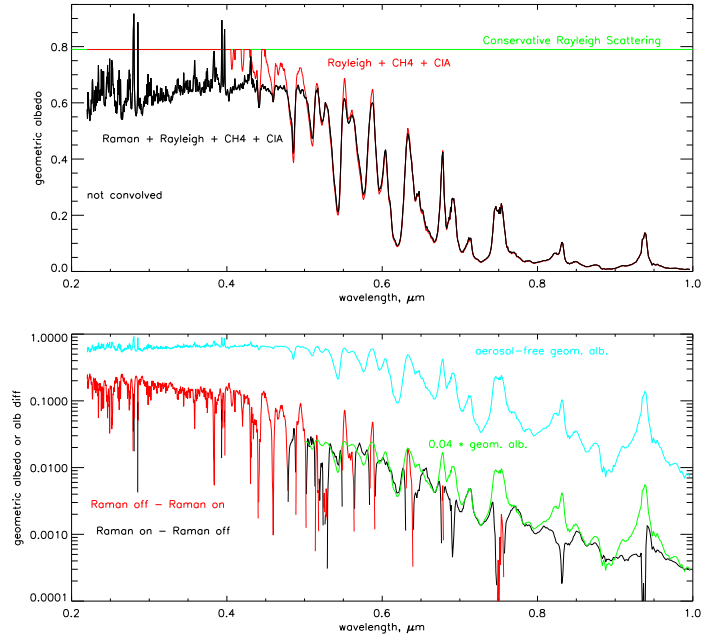}
\caption{Upper: Neptune's geometric albedo spectrum for models without aerosols assuming
conservative Rayleigh scattering with polarization (green), assuming
Rayleigh plus molecular absorption (red), and assuming Rayleigh
scattering, molecular absorption, and Raman scattering (black).
Lower: Spectral differences due to Raman scattering, with Raman losses (red) and
Raman gains (black) compared to 0.04 $\times$ the geometric albedo spectrum (green).}
\label{Fig:ramanonoff}
\end{figure}

The effect of Raman scattering on Neptune's aerosol-free geometric albedo is
illustrated in Fig.\ \ref{Fig:ramanonoff}, where spectra including
polarization are shown for three calculations of increasing
complexity, first with only Rayleigh scattering, then including also
molecular absorption, and finally including Raman scattering.  The
calculations were started at 0.2 $\mu$m, and are accurate to better
than 1\% at wavelengths beyond 0.22 $\mu$m.  Experiments with
different starting wavelengths show that an accuracy of 1\% or better
can be achieved at a given short wavelength by starting the
calculation at a corresponding wavenumber that is higher by twice the
wavenumber shift of the $Q$-branch transition. This relatively small
starting offset is effective because of the steep gradient in the
solar spectrum at short wavelengths (see Fig.\ \ref{Fig:solarspec}).
The calculation was carried out with a  step size of  17.78
cm$^{-1}$, using a solar spectrum with a nominal 
resolution of 35 cm$^{-1}$.
With only conservative Rayleigh scattering, Neptune's geometric albedo
would be a wavelength-independent 0.791 (Sromovsky 2004).  Adding
methane absorption and collision-induced hydrogen absorption (CIA)
only reduces the albedo at wavelengths beyond 0.4 $\mu$m.  Raman
scattering produces profound effects throughout the UV and most of the
visible spectrum.  

Besides the introduction of sharp spectral features, Raman scattering
also reduce Neptune's baseline geometric albedo by nearly 25\% in a
clear atmosphere.  In Fig.\
\ref{Fig:ramanonoff} the albedo decreases (shown as red) appear mostly at
shorter wavelengths, but also appear in the vicinity of most of the
local peaks in the near-IR reflectivity. The albedo gains (shown as black)
occur not just at the deep minima in the solar spectrum, but also at
the absorption maxima in the CH$_4$ spectrum, which correspond to
minima in the reflectivity spectrum.

We find that the fill-in of the reflectivity minima in the near IR is
only about 4\% of the geometric albedo.  This is in sharp contrast to
the Cochran and Trafton (1978) conclusion that in the cores of the
strong CH$_4$ bands at 0.86 $\mu$m, 0.89 $\mu$m and 1.0 $\mu$m,
nearly all of the photons that leave the atmosphere have been Raman
scattered.  
However, their conclusion is a result of what we now know to be
inappropriate assumptions concerning the amount and vertical
distribution of methane. Their standard model had a constant mixing
ratio of 0.005 relative to H$_2$ throughout the atmosphere.  This
resulted in CH$_4$ supersaturation for $p < 1$ bar,
by factors of $\sim$100 at the tropopause and more than 10 in the
stratosphere. This resulted in a very low single scattering albedo of
$\omega \sim 0.002$ at 0.89 $\mu$m, which even for a semi-infinite
atmosphere would yield a very low geometric albedo of only 0.00036,
using the single-scattering approximation
$p(\omega)=0.1855\times\omega$ for $\omega \ll 1$ (Sromovsky
2004). That is about 20 times smaller than the geometric albedo we
computed for our assumed CH$_4$ profile,
which never exceeds saturation levels and is limited to
3.5$\times$10$^{-4}$ in the stratosphere.
Our distribution leads to much more scattering at the center
of the disk and significant limb brightening (Sromovsky 2004),
while the uniform mixing ratio of Cochran and Trafton makes the atmosphere dark at all
view angles.  

\subsection{Comparison with other Raman Calculations}\label{Sec:compcal}

\begin{figure}[!htb]
\hspace{-0.35in}\includegraphics[width=3.8in]{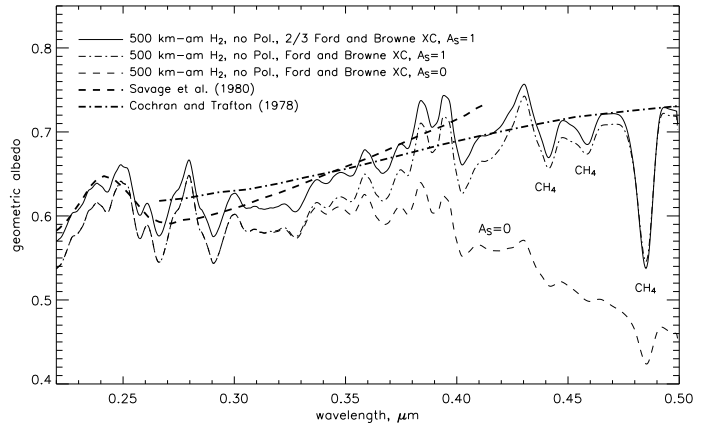}
\caption{Aerosol-free Neptune spectra computed for 500 km-amagats of H$_2$,
without He, excluding polarization, and assuming Ford and Browne (1973) Raman
cross sections (dot-dash for A$_S$=1, dashed for A$_S$=0) and 2/3 Ford
cross sections (solid), compared to earlier calculations by Cochran
and Trafton (1978)(heavy dot-dash) and by Savage \etal (1980) (heavy
dashed).  Our spectra have been smoothed to an effective resolution of
5 nm to facilitate comparisons. See text for discussion.}
\label{Fig:savage}
\end{figure}

The earliest accurate calculation of a Raman spectrum relevant to 
Neptune was by
Cochran and Trafton (1978) for a pure H$_2$ atmosphere assuming an
equilibrium population of ortho and para states. Their most relevant
case is for 500 km-amagats of H$_2$ over an opaque
cloud layer (of unspecified but apparently high albedo).  That amount
of H$_2$ corresponds to an atmospheric depth of approximately 6
bars, which is deep enough to approach semi-infinite behavior only for
$\lambda <$ 0.34 $\mu$m. At 0.5 $\mu$m the surface albedo
can make a difference of 0.25 in geometric albedo.  Using 2/3 of the
Ford and Browne (1973) Raman cross sections to approximate the values
used by Cochran and Trafton, we computed geometric albedo values for a He-free
atmosphere with a unit-albedo Lambertian surface placed at the 6-bar
level, and ignored polarization. This spectrum, smoothed to a
resolution of 5-nm, is displayed in Fig.\ \ref{Fig:savage} where it
compares well with the even lower resolution spectrum of Cochran and
Trafton.  

The other prior calculation displayed in Fig.\ \ref{Fig:savage} is due
to Savage \etal (1980), who used the same iterative computational
scheme first presented by Cochran and Trafton (1978).  According to
Savage \etal, they used the Ford and Browne (1973) Raman cross
sections that we used, but ignored polarization and assumed a pure
H$_2$ atmosphere, which was deep but of unspecified depth.  In
spite of their use of different Raman cross sections, the Savage \etal
and Cochran and Trafton results agree with each other and
 with our calculations using 2/3 of of the Ford and
Browne cross sections, raising questions about what cross sections
were actually used by Savage \etal (1980).  Our calculation with the
Ford and Browne cross sections, and other conditions being the same,
are shown as the thinner dot-dash curve in Fig.\ \ref{Fig:savage}. Clearly, this
spectrum is not compatible with the Savage \etal results. All these
calculations have a stronger upward slope than we obtained for a truly
deep atmosphere (Fig.\ \ref{Fig:ramanonoff}) because the surface
reflection becomes more visible at longer wavelengths. This can be
seen from the difference between calculations for zero and unit
surface albedos, which are shown in the figure. Other effects occur
when we carry out more realistic calculations. The presence of helium
dilutes the Raman absorption somewhat, producing about a 2\% increase
in the baseline value; using the correct Raman cross sections causes an
albedo decrease of about 0.02-0.03 (a 3-5\% decrease); and
polarization increases the geometric albedo by about 0.04 (Sromovsky
2004) (a 6.7\% increase). The approximate net result is a very small
difference between the accurate calculation with polarization and the
earlier H$_2$-only calculations without polarization.

\subsection{Comparison with Observed Spectra}\label{Sec:obscomp}

\begin{figure}[!htb]
\hspace{-0.15in}\includegraphics[width=3.7in]{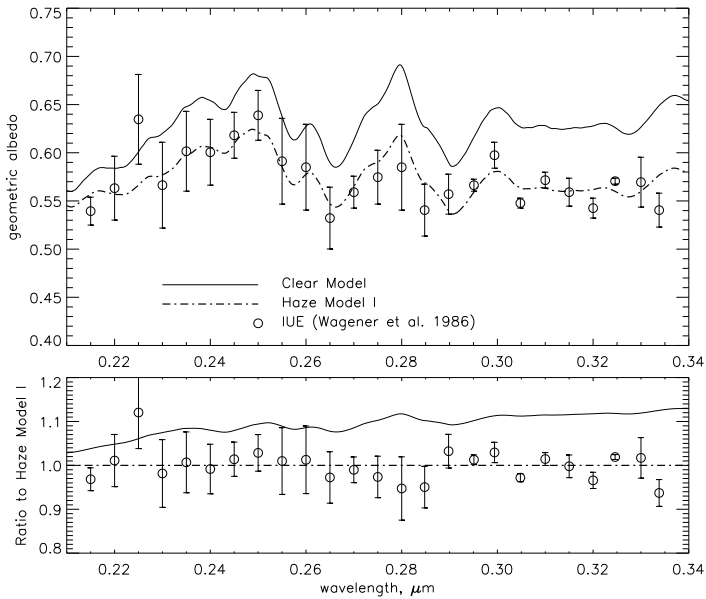}
%
\caption{Comparison of our aerosol-free (solid) and Haze I (dashed) 
calculated Neptune spectra with observations by the IUE (dot-dash). The model
spectra are smoothed to a resolution of 5-nm, to be compatible with
the IUE observations. Ratio spectra are displayed in the lower panel.}
\label{Fig:iue}
\end{figure}

The International Ultraviolet Explorer (IUE) made full-disk
observations of Neptune during 2-5 October 1985 (days 275-277) at
a phase angle of 1.87\degx and a sub-earth latitude of
-19.95\degx planetographic (-19.325\degx centric). 
Wagener \etal (1986) converted these observations to geometric albedo
assuming an equatorial radius of 25240 km and an oblateness of
0.021, which are average stellar occultation results referring to a
pressure of 1 microbar (French 1984). Using
measurements of 24764 km and 0.01708 (Davies \etal 1992), which refer
to the 1-bar level, we obtain a conversion factor of 1.035 for
adjusting the Wagener
\etal albedo values (this disagrees with the factor of 1.047 used by Courtin
(1999)). We did not make a correction for phase angle, though that
might increase their values by as much as 1\% if the phase curve at UV
wavelengths is similar to that observed at longer wavelengths
(Sromovsky \etal 2003).  The adjusted IUE results are compared with
our clear-sky and Haze I Neptune spectra in Fig.\
\ref{Fig:iue}, where error bars indicate the absolute uncertainty estimates
due to Wagener \etal (1986). The IUE results are  5\% to 10\% below the theoretical
 clear-sky calculated spectrum, generally showing the smaller
 difference at shorter wavelengths (0.21-0.23 $\mu$m) and the best
 defined and larger differences at longer wavelengths (0.29-0.33
 $\mu$m), which is likely due to an absorbing
 high altitude haze.  Our Haze I model calculations match the IUE
 observations generally well within the IUE error bars.  If we had
 neglected polarization, the model calculations would drop about 5\%,
 making the clear-atmosphere model the best fit near 0.25 $\mu$m.  In
 this spectral range, polarization makes the difference between
 needing haze absorption and not needing it.

\begin{figure}[!htb]
\hspace{-0.25in}\includegraphics[width=3.7in]{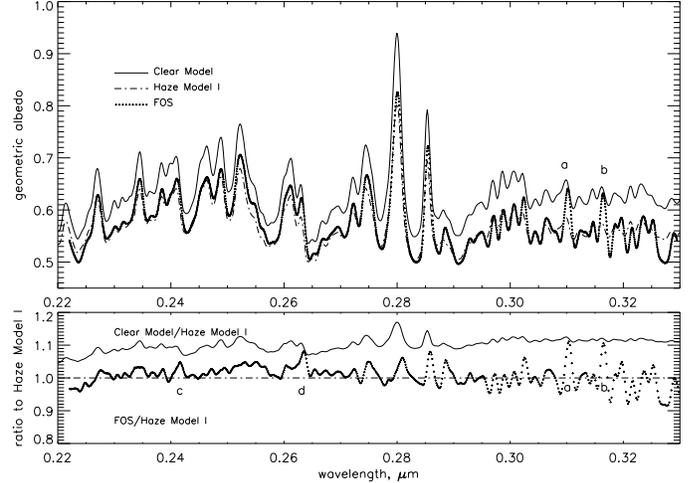}
%
\caption{Comparison of our aerosol-free (dot-dash) and Haze I (solid)
calculated Neptune spectra with FOS observations (+++) by Courtin (1999). 
All spectra are convolved to a spectral resolution of 1-nm.}
\label{Fig:fos}
\end{figure}

The Faint Object Spectrograph (FOS) observed Neptune at a phase angle
of 1.22\deg and a sub-earth latitude of -24.78\deg planetographic on
19 August 1992 using apertures of 0.3$''$, covering 21-36\degx S, and
1.0$''$, covering 3-35\degx S. Because these apertures are smaller
than Neptune's 2.4$''$ disk and placed near the middle of Neptune's
disk, the observed I/F and the relative amplitudes of Raman spectral
features are increased relative to what would be observed for a
full-disk average (shown in Sec.\ \ref{Sec:angle}).
Courtin (1999) obtained uncalibrated geometric
albedo spectra by dividing FOS flux spectra by a solar flux spectrum
measured by the UARS SOLSTICE instrument (Rottman \etal 1993; Woods
\etal 1993).
The FOS results we use are the 1-arcsec spectra degraded to a
resolution of 1 nm to improve the signal/noise ratio (shown in
Courtin's Fig. 2).  Courtin performed a radiometric calibration by
matching his 0.22-0.33 $\mu$m FOS spectra to full-disk 0.3 - 1.0
$\mu$m groundbased spectrum of Karkoschka (1994), obtained during
23-26 July 1993 at a phase angle of 0.4\degx.  The FOS observations
thus have the same 4\% absolute uncertainty as the Karkoschka (1994)
reference spectrum. The FOS noise level varies from 0.4\% at 0.33
$\mu$m to 1\% at 0.26 $\mu$m.
Courtin's Neptune spectrum is compared to our clear-sky and Haze I
calculations in Fig.\ \ref{Fig:fos}. Our calculations as a function of
view angle were integrated over approximately the same field of view
as the FOS observations to account for the increased Raman effect near
normal view angles.
We then convolved them to a final resolution of 1 nm and scaled this
average spectrum to match our calculated geometric albedo spectrum in
the wavelength range from 0.31 $\mu$m to 0.33 $\mu$m, to simulate the
calibration procedure used by Courtin.  Note the excellent agreement
between the shapes and amplitudes of the calculated spectral features
for the Haze I model and those measured by the FOS.  The most glaring
exceptions (a and b in the figure) appear to be artifacts in the FOS
spectrum.  These anomalies are not present in the overlapping portion
of the Karkoschka spectrum, which is in good agreement with the
calculated spectrum.  The spikes c and d in the ratio spectrum are at
points where the I/F spectrum has very sharp gradients, and could be
removed by a slight wavelength shift.  The FOS/Haze I ratio spectrum
generally agrees with the IUE/Haze I ratio shown in Fig.\
\ref{Fig:iue}, indicating that haze absorption may be depressing the
reflectivity by about 6\%-10\% below the theoretical value for a clear
Neptune atmosphere in the 0.23-0.27 $\mu$m range, with increasing
absorption indicated at longer wavelengths. 

\begin{figure}[!htb]
\hspace{-0.2in}\includegraphics[width=3.7in]{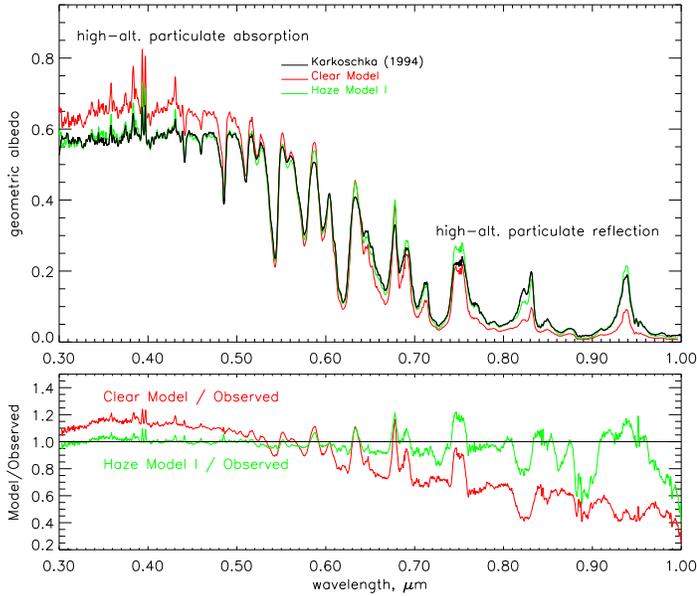}
%
\caption{Comparison of calculated aerosol-free (red) and
Haze I (green) model geometric albedo 
spectra of Neptune with groundbased observations (black line) 
by Karkoschka (1994). The ratios of model calculations to
observed are shown in the lower panel.}
\label{Fig:karkcompto1}
\end{figure}

Figure\ \ref{Fig:karkcompto1} compares clear-atmosphere and Haze I
model calculations with the disk-integrated groundbased spectra of
Karkoschka (1994). Between 0.35 $\mu$m and 0.45 $\mu$m the
observations are about 13\% below the clear-atmosphere calculations,
indicating a presence of an absorbing haze that is even more
influential than at shorter wavelengths. This behavior suggests that
the haze extends deep enough into the atmosphere that it becomes more
influential as the molecular scattering optical depth above it
decreases with increasing wavelength. This is qualitatively consistent
with microphysical models of Moses \etal (1995), which suggest that
haze opacity increases down to the 870 mb level. It is also roughly
the character of our Haze I model, which is shown to provide good
agreement with Karkoschka's observations, although the computed
amplitudes of the Raman spectral features are somewhat larger than the
observations in several cases.  This discrepancy might mean that more
aerosol opacity is needed in the 1-3 bar level, or that there is a
slight difference in the effective spectral resolution of the
observations and the calculations.  At longer wavelengths, the high
observed albedo values in the deep CH$_4$ bands at 0.8 $\mu$m and 0.9
$\mu$m, indicate that there is also particulate scattering at high
altitudes, which is not included in either of the haze models.  Much
of that contribution probably comes from discrete bright features seen
in Voyager, HST, and groundbased images. Effective pressure estimates
for such features range from 23 mb to 60 mb at 30-40\deg N, 100-230 mb
at 30-50\deg S, and 170 mb to 270 mb near 70\deg S (Sromovsky \etal
2001b; Gibbard \etal 2003).  The effective fractional coverage of
bright high altitude clouds required to explain the albedo in deep
CH$_4$ bands appears to be about 0.5\% to 1\% for an upper cloud at
150 mb (Sromovsky \etal 2001a). The effective fractional coverage of
a coexisting 1.3-bar cloud is $\sim$1\% assuming a unit-albedo
Lambertian reflector.

The observed albedo in the window regions at 0.825 $\mu$m and
0.935 $\mu$m is also significantly above the clear-sky calculation,
suggesting a significant aerosol contribution that could be at mid to
deep levels, which is largely satisfied by the 1.3-bar and 3.8-bar
clouds in the Haze I model.  The excess I/F calculated in the window
regions at 0.59 $\mu$m, 0.63 $\mu$m, 0.68 $\mu$m, and 0.75 $\mu$m, can
be substantially reduced by increasing the CH$_4$ continuum
absorption by amounts comparable to Karkoschka's stated continuum
uncertainty.  It could also be reduced by decreasing the
single-scattering albedo of the cloud aerosols.  The observations seem
to show much less of the collision-induced absorption in the 0.8-0.83
$\mu$m region than is evident in the calculations. Another way to
describe this difference is to say that there is an imbalance in the
0.83 and 0.94 $\mu$m window regions that is evident in the
calculation, but not in the observations.  That imbalance may be due
to an error in the CH$_4$ absorption coefficients.  If we add to
Karkoschka's standard coefficients the continuum absorption
uncertainty, we actually obtain very similar albedo values in these
two windows.

\subsection{Spectral Evidence of the Ortho/Para Ratio}

\begin{figure}[!htb]
\hspace{-0.2in}\includegraphics[width=3.7in]{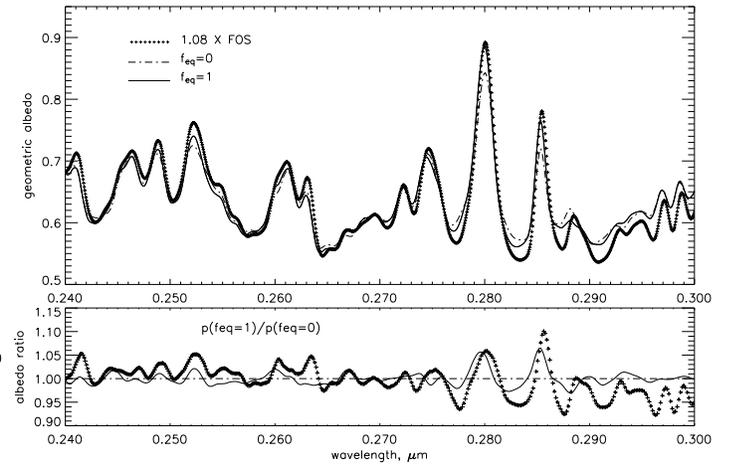}
%
\caption{Geometric albedo spectra for equilibrium H$_2$ (solid) and
normal H$_2$ (dot-dash) compared to the observed FOS spectrum (plus
signs) (Courtin 1999) scaled by the factor 1.08.  The ratio of
equilibrium to normal albedos is in the lower panel.
}
\label{Fig:feqcomp}
\end{figure}

Because the $J=0$ Raman cross sections are larger than the $J=1$ cross
sections (Fig.\ \ref{Fig:xcvswlen}), the ortho/para ratio will have an
effect on the Raman features observed in Neptune's reflection
spectrum.  The largest difference in ortho/para ratios for equilibrium
and normal H$_2$ occur near 100 mb (Fig.\ \ref{Fig:parapop}), and thus
associated spectral variations are more likely at short wavelengths
where this region is near the $\tau=1$ level (Fig.\
\ref{Fig:pendepth}).  Calculations (Fig.\
\ref{Fig:feqcomp}) show that the largest difference is for the peaks
 at 0.280 $\mu$m and 0.2852 $\mu$m, which are about 5\% larger for
 equilibrium H$_2$.  The overlay of FOS observations suggests better
 agreement with equilibrium H$_2$, as concluded by Courtin (1999),
 although the relative size of the smaller peak is very dependent on
 the effective resolution of the observations and both peaks have
 amplitudes that depend on view angle.  And since the FOS observations
 cover only the central disk, it is not strictly valid to compare them
 to geometric albedo observations.  Uncertainties in the wavelength
 scale and effective resolution of the FOS observations arise from
 comparisons with overlapping groundbased observations of Karkoschka
 (1994), which show significantly less spectral modulation in the
 0.3-0.33 $\mu$m overlap region, as noted earlier.  The presence of an
 absorbing haze and high-altitude discrete clouds also affects these
 peaks.  Thus it seems premature to make a strong conclusion based on
 these observations.  However, Courtin's (1999) result that
 $f_{eq}$=0.88$\pm$0.23 is based on higher resolution observations
 than those shown in Fig.\ \ref{Fig:feqcomp}, and was derived after
 special processing to minimize wavelength errors.  Courtin's results
 are also compatible with independent conclusions by Baines and Smith
 (1990) and Conrath \etal (1991) that the ortho/para ratio for Neptune
 is near thermal equilibrium.

\section{Approximations of Raman Scattering}

\subsection{The Karkoschka Correction}\label{Sec:kcorrect}

Karkoschka (1994) presented a method for transforming spectra between
Raman and non-Raman forms 
based on the assumption that the
 measured geometric albedo spectrum $p(\nu)$ can be well approximated
by a linear combination of terms involving offset versions of the
``physical'' spectrum $q(\nu)$, which is the reflectivity spectrum without
Raman scattering.  The mathematical expression of this method is
\begin{eqnarray} p(\nu)=f_0(\nu)q(\nu) + \nonumber \\
\sum_{\Delta\nu}f_{\Delta \nu}(\nu+\Delta\nu)q(\nu+\Delta\nu)
F_\odot(\nu+\Delta\nu)/F_\odot(\nu)
\label{Eq:karkcorr}\end{eqnarray}
where $F_\odot$ is here the solar photon spectrum, $f_0(\nu)$ is the
fraction of photons not Raman scattered, $f_{\Delta
\nu}$ is the fraction undergoing transition $\Delta \nu$, and where
$f_0(\nu)+\Sigma f_{\Delta \nu}(\nu)=1.$ Equation\ \ref{Eq:karkcorr}
can be inverted to obtain the Raman-free ``physical'' spectrum from
the observed spectrum using $q_1(\nu)=p(\nu)$ as a first guess, and
then improving the guess using the iterative equation
\begin{eqnarray} q_{n+1}(\nu) = f_0(\nu)^{-1}\big[ p(\nu)
 - \nonumber \\ \sum_{\Delta\nu}f_{\Delta \nu}(\nu+\Delta\nu)q_n(\nu+\Delta\nu)
 F_\odot(\nu+\Delta\nu)/F_\odot(\nu)\big]
 \label{Eq:karkinv},\end{eqnarray} 
for which four iterations are usually sufficient to converge on $q$
for a given set of fractions.  Karkoschka assumed a power law
$\lambda$-dependence for $f_{\Delta \nu}$, then used minimum roughness
of the fitted spectrum for 0.31 $\mu$m $< \lambda <$ 0.405 $\mu$m as
the criterion for picking the optimum fractions and exponent.

\begin{figure*}[!t]\centering
\includegraphics[width=6in]{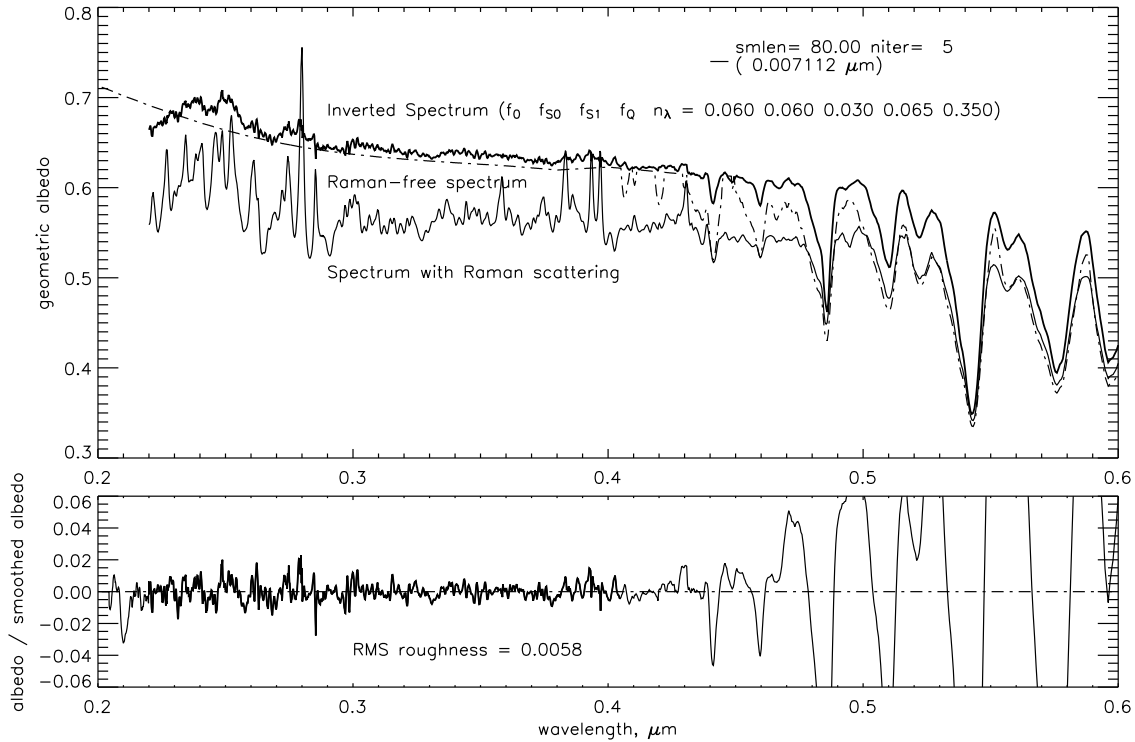}
%
\caption{Application of Karkoschka's method of removing effects
of Raman scattering from the model spectrum (thin solid line) for our haze/cloud
aerosol structure. The inverted ``physical'' spectrum (thick solid line) is
slightly above the true Raman-free spectrum (dot-dash line). The smoothing
length used in judging fit quality was 0.71 nm. Lower: the ratio of the fit to the
smoothed fit, with the heavier portion of the curve indicating the part used
to constrain the fit.}
\label{Fig:ramremove}
\end{figure*}

\begin{table*}\centering
\caption{Direct Fits of Karkoschka's Empirical Model to Sample Spectra.}
\begin{tabular}{c c c c c c c c}
\hline
\rule{0pt}{0.15in}
      & $\lambda$-range & & & & & & RMS \\ Spectrum & ($\mu$m) & $f_0$
      & $f_{S0}$ & $f_{S1}$ & $f_Q$ & $n_\lambda$ & DEV \\
\hline
\rule{0pt}{0.15in}
 Clear Sky & 
   0.31-.405  & 0.100$\pm$0.005 & 0.065$\pm$0.007 & 0.050$\pm$0.005 & 0.085$\pm$0.005 & -0.80$\pm$0.05 & 0.93\% \\
 Clear Sky 
 & 0.23-.405  & 0.100$\pm$0.005 & 0.070$\pm$0.005 & 0.030$\pm$0.005 & 0.091$\pm$0.005 & -0.50$\pm$0.07 & 1.47\% \\
 Haze II& 
   0.31-.405  &0.050$\pm$0.005 & 0.065$\pm$0.005 & 0.030$\pm$0.005 &0.057$\pm$0.005 & -0.55$\pm$0.05 & 0.44\% \\
 Haze II&  
 0.23-.405  &0.035$\pm$0.003 & 0.060$\pm$0.004 & 0.030$\pm$0.005 & 0.085$\pm$0.005 & -0.35$\pm$0.06 & 0.90\% \\
 Haze II (inv)&  
 0.23-.405  &0.060$\pm$0.020 & 0.060$\pm$0.004 & 0.030$\pm$0.003 & 0.065$\pm$0.005 & -0.35$\pm$0.05 & 0.58\% \\
 K94 Coeff.&  0.31-.405  
      & 0.035$\pm$0.015 & 0.060$\pm$0.01 & 0.025$\pm$0.015 & 0.050$\pm0.015$ & -1$\pm$1 &\\[0.5ex]
\hline
\end{tabular}
\label{Tbl:fracfits}
\parbox{5.5in}{NOTE:  Fractional
values are given at 400 nm. Values at other wavelengths are
proportional to $\lambda^{n_\lambda}$. Haze II (inv) refers to the
inversion of fit parameters from the Raman spectrum only, rather than
starting with the Raman-free spectrum. In this case RMS DEV is
relative to the smoothed spectrum, which was smoothed with
a boxcar of 0.71 nm. The last row is from Karkoschka
(1994).}
\end{table*}

We first investigated how well the proposed model could fit Raman
spectral calculations and how the fitted constants related to Raman
scattering cross sections. We considered both a clear-atmosphere model
and a model with haze and cloud contributions.  We first fit the
fractional parameters and wavelength dependence exponent by minimizing
the RMS deviation between the Karkoschka model and the calculated
spectrum, first over the 0.31-0.405 $\mu$m range used by Karkoschka
and also over the wider 0.23-0.405 $\mu$m range.  The results are
summarized in Table\ 5. Note that our fit results are comparable to
those of Karkoschka, especially when applied to a hazy atmosphere.
Although we were able to obtain good fits over the limited spectral
ranges (RMS deviations of 0.44\% and 0.9\% for the haze/cloud model,
and 0.93\% and 1.47\% for clear sky model), the fits did not
accurately match the observed spectrum at slightly longer wavelengths,
especially where CH$_4$ absorption was present.  The weak CH$_4$
bands in the Karkoschka conversion of the physical spectrum
were much stronger than the corresponding bands in the correct Raman
calculation.  This is understandable as a consequence of the nonlinear
relationship between I/F and single-scattering albedo.  A small
absorption added to a conservative atmosphere can have a much larger
fractional effect than it does when added to an absorbing atmosphere.
For example, adding $k_{abs} = 0.003 \times k_{scat}$ to a Rayleigh
atmosphere with $k_{abs}=0$ changes $\omega$ from 1.0 to 0.997, which
decreases the I/F from 0.791 to 0.699, a 9\% drop.  But if the I/F is
already at 0.559 due to Raman absorption, adding the same CH$_4$
absorption as before would change $\omega$ from 0.975 to 0.972,
producing only a 2\% drop in I/F.
Confirming Karkoschka's fit results, we find that the vibrational coefficient
($f_Q$) is about twice the size of the coefficient for the $J=1$
rotational contribution ($f_{S1}$), even though its Raman cross
section is somewhat smaller (Table\ 3).  Since we use these cross sections
to compute the spectra, the fit results do not suggest that there
is something wrong with the cross sections. A large part
of this result comes from the fact that the vibrational coefficient
represents the average of ortho and para contributions, while the
individual $S(0)$ and $S(1)$ contributions are each multiplied by
fractions that decrease their relative contributions by roughly a
factor of two (for $f_{para}\approx 1/2$, see Eq.\ \ref{Eq:sigtot} and
Fig.\ \ref{Fig:parapop}). Additional factors are discussed in Sec.\
\ref{Sec:monoscat}.

We next consider how well Karkoschka's method is able to remove Raman
scattering effects from a spectrum that includes Raman scattering. We
inverted our model Raman calculations using the inversion method
based on Eq.\ \ref{Eq:karkinv}.
The retrieved parameter values for the Haze II test case
 are given in Table\ 5. 
As illustrated in Fig.\ \ref{Fig:ramremove}, we found pretty fair
agreement between the true Raman-free spectrum and inverted spectrum
for the haze/cloud case when we used the 0.23-0.405 $\mu$m range for
the fitting, although our inverted spectrum is offset about 0.01
albedo units above the true spectrum, and the inverted parameters
differ somewhat from the best-fit parameters.  Using the 0.23-0.405 $\mu$m 
spectral range yielded a smoother inverted spectrum but a
larger offset albedo of 0.03.  Much worse results were obtained
with the clear sky model, which has larger Raman features that are not
as well fit by Karkoschka's empirical model.  As expected, we found
that the features due to weak CH$_4$ bands were much smaller in the
transformed spectra than in the true physical spectra.  Thus,
estimates of CH$_4$ absorption coefficients using spectra corrected
in this way will be significantly below the true absorption level.

Karkoschka's method has several physical flaws
that limit its applicability.  Multiple scattering is one effect
that is not accounted for but can be significant (see Sec.\ \ref{Sec:monoscat}). 
Another problem with the physics assumed in
Eq. \ref{Eq:karkcorr} is that the contribution
at the present wavenumber is not necessarily proportional to the geometric
albedo at the upward shifted wavenumber.  
In fact, the proportionality constant also depends on the vertical
distribution of the scattered photons in relation to the vertical
distribution of opacity at the scattered wavelength. 
The Karkoschka forward method also is not generally applicable to spatially
resolved observations because the amplitude of Raman effects
depends on view angle as well as local aerosol structure.
The useful spectral range for the Karkoschka correction is also rather
limited, probably to wavelengths less than about 0.405 $\mu$m, partly because
it does not properly modify weak CH$_4$ absorption bands.

\subsection{The Wallace Approximation}

\begin{figure}[!htb]
\hspace{-0.2in}\includegraphics[width=3.7in]{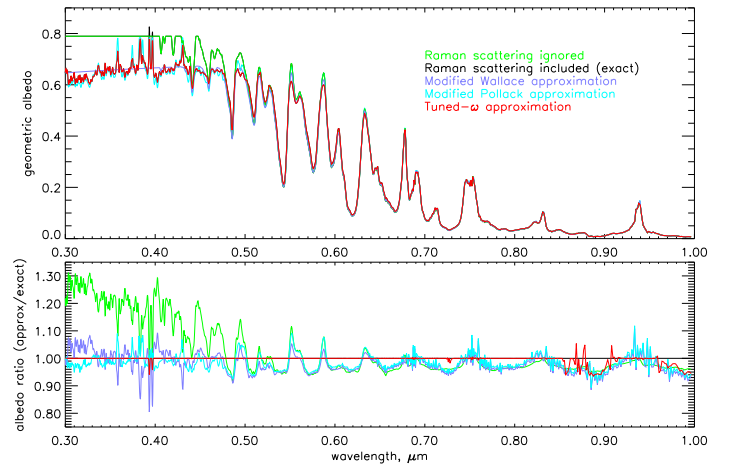}
\caption{Upper: Geometric albedo spectrum computed for a clear Neptune atmosphere
with no accounting for Raman scattering (blue), with Raman scattering
included (black), and with Raman scattering approximately accounted
for using a modification of the Wallace (1972) approximation (violet),
a modified and generalized version of the Pollack (1986) approximation
(blue) and the tuned-$\omega$ approximation (red). Lower: Ratios of
the approximate geometric albedo spectra to the accurate Raman
calculation.}
\label{Fig:approx3in1}
\end{figure}

The approximation suggested by Wallace (1972) is a modification of the molecular single
scattering albedo of the atmosphere by treating rotational
transitions as extra sources of scattering and the vibrational transition
as a pure absorption. This can be expressed in terms of cross sections as 
follows: \begin{equation}
 \omega_{W} = \frac{\sigma_{\mathrm{Ray}} +  \sigma_{\mathrm{rot}} +
   \beta \sigma_{\mathrm{vib}}}
{\sigma_{\mathrm{Ray}} + \sigma_{\mathrm{rot}} + \sigma_{\mathrm{vib}}
+ \sigma_{\mathrm{abs}}}, \label{Eq:wallace}
\end{equation}
where $\sigma_\mathrm{Ray}$ is the Rayleigh scattering cross section,
$\sigma_\mathrm{abs}$ is the total cross section for absorption by
CH$_4$ and CIA by H$_2$,
$\sigma_\mathrm{rot}$ is the combined cross section for rotational
Raman transitions, $\sigma_\mathrm{vib}$ is the cross section for
vibrational Raman transitions, and where $\beta=0$ for the original Wallace
approximation.  For a clear Neptune
atmosphere,
we find that the original Wallace approximation provides a crude match
to the low resolution baseline of the spectrum, but is generally 5\%
to 10\% low and of course does not produce characteristic Raman
spectral features.  A few detailed calculations by Wallace (1972) at
0.2 $\mu$m and 0.4 $\mu$m also indicated that his approximation was
about 5\% low.  The Wallace approximation error is nearly zero in the
near-IR CH$_4$ windows, but is about 4\% low in the CH$_4$ absorption
bands, which is about what would be expected if Raman scattering were
ignored (see lower part of Fig.\
\ref{Fig:approx3in1}).  Figure\ \ref{Fig:approx3in1} displays a modified
form of the Wallace approximation that boosts the I/F value by setting
$\beta = 0.433$, which treats the vibrational cross section as 56.7\%
absorption and 43.3\% scattering.  This improves the overall agreement
in the blue to orange part of the spectrum, but has little effect
elsewhere.  The Wallace approximation approaches the exact result when
high clouds obscure most of the molecular scattering.

\subsection{The Pollack Approximation}

Pollack \etal (1986) approximated Raman scattering contributions from
shorter wavelengths by scaling the Raman scattering cross section at
the reflected wavelength by the ratio of solar irradiance at the source
wavelength to that at the reflected wavelength. Generalizing this
approach to include more than one transition, the molecular
single-scattering albedo in this approximation can be written as:
\begin{equation}
\omega_{P}(\nu) = \mathrm{min}\Big(
\frac{\sigma_{\mathrm{Ray}}(\nu)
+ \sum_{\ell} \sigma_{\mathrm{R}\ell}(\nu+\Delta\nu_\ell)
g_\ell (r_\ell)}
{\sigma_{\mathrm{Ray}}(\nu) +\sigma_{\mathrm{abs}}(\nu)
 + \sum_{\ell} \sigma_{\mathrm{R}\ell}(\nu)},1\Big) \label{Eq:pollack}
\end{equation}
where $r_\ell=F_\odot(\nu+\Delta\nu_\ell)/F_\odot(\nu)$, $g_\ell$ is a
function of $r_\ell$ that can be empirically adjusted to compensate
for some of the complexities that are glossed over by this concept,
and the upper bound of 1 is used to avoid energy conservation
problems. Setting $g_\mathrm{vib}(r_\mathrm{vib})=r_\mathrm{vib}$, and
omitting the rotational transitions completely, would reduce this to
Pollack's original suggestion, while setting $g_\mathrm{rot}=1$ and
$g_\mathrm{vib}=\beta$ would reduce this to the Wallace
approximation. A problem with Pollack's original formulation was that
it resulted in $\omega_P > 1$ at some wavelengths, as noted by Courtin
(1999), which leads to instability in the solution of the radiation
transfer equation. Another problem is that it considered only the
$Q_1$ vibrational transition, which alone would be unable to generate
the ghost features arising from the rotational transitions.  The form
given in Eq.\ \ref{Eq:pollack} has the capability to fit all the main
ghost features and has the potential, with tuning of the form of
$r_\ell$, to yield the correct baseline value as well.  It also has an
advantage over Karkoschka's correction, in that it is introduced at a
more fundamental physical level that is naturally modified by the
presence of aerosols.  However, wherever the geometric albedo exceeds
the value for a conservative Rayleigh atmosphere of 0.7908 (Sromovsky
2004), even $\omega_P =1$ will not be able to reproduce the albedo
value. This problem can be reduced by working at lower spectral
resolutions for which these extreme peak amplitudes become reduced
below 0.7908.

Preliminary calculations with this approximation used the three
transitions given in Table\ 4, with $g_\ell =
c_\ell r_\ell$ and $c_\ell$ the same for all transitions.  We found
that making the constant $c_\ell$ = 0.875 for all $\ell$ produced a
reasonably good match to the baseline level of geometric albedo, but
generated spectral modulations that were too large. To limit these
modulations, we used an exponential function to limit the effect of
the spectral irradiance ratio, i.e. we used $g_\ell =
w(1-\exp(-r_\ell/w))$, which is $\sim r_\ell$ for small $r_\ell$ and
approaches $w$ for large values of $r_\ell$.  We found that $w=2$
provided a reasonable fit to the Raman spectral features and to the
baseline level, as shown in Fig.\
\ref{Fig:approx3in1}.  While this is certainly an improvement over the
Wallace approximation, and does reproduce many of the spectral
features associated with Raman scattering, it does not do so
consistently and accurately enough to use for interpreting those features.  Like
the Wallace approximation, it also fails to improve geometric albedo
accuracy for $\lambda >$ 0.47 $\mu$m.  Both Wallace and
Pollack modifications to the single-scattering albedo seem to be far
less effective in regions where CH$_4$ absorption is important. Since
the actual source function depends on the light level in the atmosphere,
which may have a spectral shape that is very different from the incident
solar spectrum, it is not appropriate to use the solar spectral
ratio as a scale factor in spectral regions with strong CH$_4$ absorption.
The following method goes part of the way in solving this problem.

\subsection{Approximation of Raman Scattering by Spectral Tuning of $\omega$}

The following describes a new approximation of Raman scattering that
uses a different kind of modification of the molecular single
scattering albedo.  Instead of using a constant factor or the solar
spectral ratio to modify the molecular single-scattering albedo, this
approximation uses a $\lambda$-dependent multiplier that is tuned to
minimize the difference between simulated and exact Raman calculations
of geometric albedo for an appropriate model atmosphere.
The fundamental relation used is from Sromovsky (2004):
\begin{equation} p(\omega)= 0.7908 \omega^{-0.269}
\big[1-\exp(-|\log_{10}(1-\omega)|^{1.269}/1.504)\big], \label{Eq:pofomega}
\end{equation}
where $p$ is the geometric albedo including polarization contributions
and $\omega$ is the single scattering albedo of a semi-infinite
Rayleigh atmosphere. This equation is inverted numerically to provide
a function $\omega(p)$ so that a given geometric albedo can be related
to an equivalent single-scattering albedo of a semi-infinite Rayleigh
atmosphere.  Starting with a Raman-free calculation producing a
spectrum $p_{NR}(\nu)$, and a true Raman calculation producing a spectrum
$p_{R}(\nu)$, we use the inverse transform to compute equivalent
single-scattering albedos $\omega_{p,NR}$ and
$\omega_{p,R}$.  Noting that $1/\omega = 1 + k_{abs}/k_{scatt}$, we then
attribute the difference between the equivalent inverse single scattering
albedos as due to a difference in the effective absorption due to Raman
scattering.  Thus we are led to the approximation for modifying the
single scattering albedo of the molecular scatterings that simulate true Raman
scattering, namely, \begin{equation}
 \omega_{sim,n+1} = \Big[ 1/\omega_{sim,n} - 
            (1/\omega_{p,n} -1/\omega_{p,R}) \Big]^{-1}
\label{Eq:tuneit1}
\end{equation}
where $\omega_{sim,n}$ is the single-scattering albedo simulation of
Raman scattering on the prior iteration, $\omega_{p,n}$ is the
effective semi-infinite single scattering albedo obtained from the
spectrum generated using $\omega_{sim,n}$ for the molecular scattering.
Because Neptune's atmosphere has a vertically inhomogeneous CH$_4$
distribution, and thus will not satisfy
Eq. \ref{Eq:pofomega}, it is necessary to iterate a few times to
achieve convergence.  On the first iteration, we have
$\omega_{sim,n=1}=1.0$ and $\omega_{p,n=1} =\omega_{p,NR}$.  
The molecular single-scattering albedo spectrum that results is $\omega_{T}(\nu) =
 \frac{\omega_{sim}(\nu) \sigma_{\mathrm{Ray}}(\nu)}
 {\sigma_{\mathrm{Ray}}(\nu) +\sigma_{\mathrm{abs}}(\nu)}$, 
which reduces to $\omega_T = \omega_{sim}$ when $\sigma_{abs}=0$.

The nearly perfect agreement between the geometric albedo computed
with Raman scattering and the approximation using a tuned
single-scattering albedo is shown in Fig.\ \ref{Fig:approx3in1}.
Unlike the previously discussed approximations, it does provide useful
accuracy for $\lambda >$ 0.45 $\mu$m.  Like the other
approximations, it cannot reproduce the peaks that exceed the I/F for
a unit geometric albedo.  While the modified single-scattering albedo
can exceed unity, or even go negative, that does not lead to problems
as long as the other contributors to the scattering process lead to a
combined single-scattering albedo $\le 1$.

The important question is whether this approximation or any of the
others that are similarly constructed, are useful with clouds present,
and for center-to-limb scans.  Because all of these approximations
that modify single-scattering albedos really lack the essential
physics of the Raman scattering process, it is doubtful whether they
can have much utility in situations for which they have not been
tuned.  A partial test of this utility is the degree to which they can
simulate the angular variation of Raman scattering, which is described
in the next section. Section\ \ref{Sec:aereffects} compares their
performances when haze and cloud aerosols are present.

\section{Angular Dependence of Raman Scattering Effects}\label{Sec:angle}

\begin{figure}[!htb]\centering
\hspace{-0.2in}\includegraphics[width=3.6in]{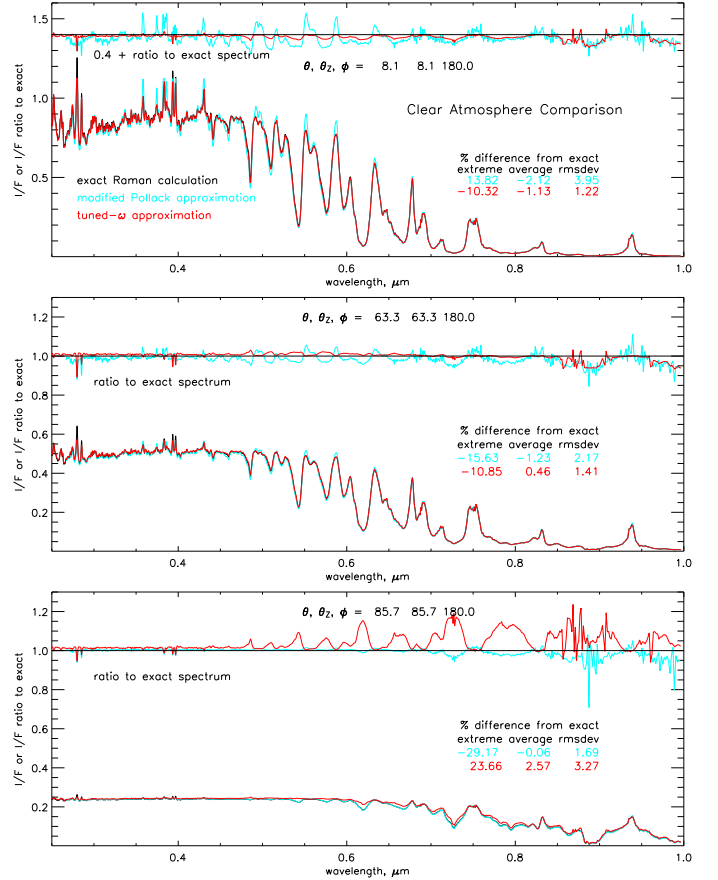}
\caption{Comparison of Neptune's I/F spectra computed with Raman scattering
 (black) and with a modified Pollack approximation (blue) and
 tuned-$\omega$ approximation (red), for three observing angles at
 zero phase. The ratios of the two spectral versions are shown at the
 upper part of each panel; in the top panel the ratio is offset by 0.4
 to prevent overlap.  The spectra are convolved to a resolution of 36
 cm$^{-1}$.}
\label{Fig:angfix2in1clr}
\end{figure}

Raman scattering effects on Neptune's I/F spectrum at zero phase angle
are strongest near the center of the disk (zero zenith angle) and
weakest near the limb (Fig.\ \ref{Fig:angfix2in1clr}).  
The errors obtained by
ignoring Raman scattering (not shown) are quite substantial at moderate
view angles, especially at short wavelengths and at high spectral resolution,
ranging from about 20 to 60\% in the 0.25-0.3 $\mu$m range at a zenith
angle of 8.1\degx. At view angles of $\sim$60\degx, the fractional errors are
comparable to those seen for the geometric albedo.  As noted for the geometric
albedo, errors $\sim$4\% are seen in the deep CH$_4$ bands, with very little
error seen in the near-IR window regions.  In the longer wavelength bands, the
fractional error does not improve much near the limb, unlike the error at 
shorter wavelengths, which decreases substantially.

\begin{figure}[!htb]\centering
\hspace{-0.15in}\includegraphics[width=3.5in]{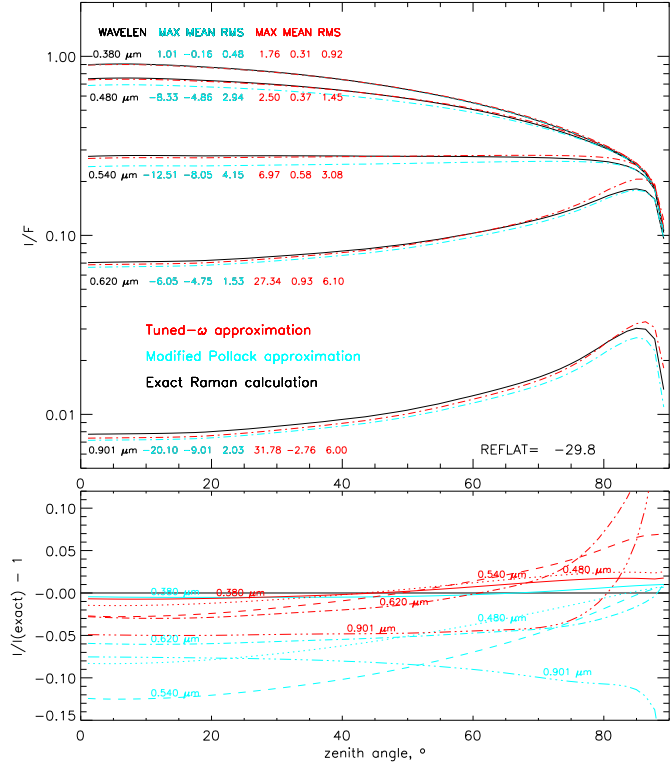}
\caption{Comparison of Neptune's center-to-limb profiles at 5 selected
wavelengths computed with Raman scattering (solid) with a modified
Pollack approximation (blue) and tuned-$\omega$ approximation (red).  
The ratios of the two spectral
versions are shown in the lower panel.  The calculations were convolved to a
resolution of 36 cm$^{-1}$ before sampling the selected wavelengths.}
\label{Fig:ctl2in1}
\end{figure}

The modified Pollack approximation helps considerably in reducing
errors at short wavelengths (Fig.\ \ref{Fig:angfix2in1clr}). The
extreme error near the central disk drops from 62\% to 14\% and the
RMS error from 19\% to 4\%. At 63\deg the RMS error drops from 8.8\%
to 2.2\%.  But at long wavelengths, the Pollack approximation actually
makes things slightly worse in the middle of strong absorption bands,
probably because the vertical location of Raman photons becomes more
critical to the resulting I/F level. Overall, the modified Pollack
approximation has value in modeling a variety of observations.  It is
far better than ignoring Raman scattering and is roughly twice as
accurate as the modified Wallace approximation. However, using the
modified Pollack approximation can lead to substantial errors in
limb-darkening profiles, depending on the wavelength that is
considered.  Sample profiles are shown in Fig.\
\ref{Fig:ctl2in1}.

Under most conditions, the tuned-$\omega$ approximation is even more accurate than the modified
Pollack approximation, as evident from comparison of spectra at the three sample angles (Fig.\
\ref{Fig:angfix2in1clr}) and from a comparison of center-to-limb scans at
sample wavelengths (Fig.\ \ref{Fig:ctl2in1}) .  The extreme error near the central disk drops
to 10.3\% and the mean and RMS deviations to only -1.1\% and 1.2\%,
with comparable results at 63\deg view angles.  The worst errors are
seen at large view angles. At 85.7\deg, the extreme error increases to
24\% and the mean and RMS errors increase to 2.6\% and 3.3\%.  But
most of these errors occur in the CH$_4$ bands beyond about 0.53
$\mu$m.  If restricted to 70\deg view angles and $\lambda <$
0.85 $\mu$m, this approximation is remarkably effective, with errors
rarely exceeding a few percent.


\section{Effects of Aerosols on Approximation Performance}\label{Sec:aereffects}

\begin{figure}[!htb]
\hspace{-0.2in}\includegraphics[width=3.7in]{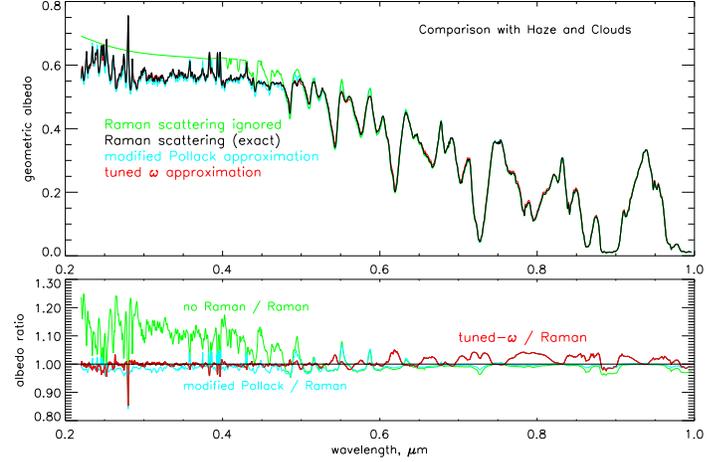}
\caption{Geometric albedo of Neptune for a sample aerosol structure
containing an absorbing high-altitude haze and two layers of clouds
(Haze II model, given in Sec.\ \ref{Sec:cloudandhaze}), computed with
Raman scattering ignored (green), using the Pollack approximation
(blue), the tuned-$\omega$ approximation (red), and the exact method
of calculation (black).  The ratios to the true calculation are
displayed in the lower panel.}
\label{Fig:approxcomp}
\end{figure}

Calculations for the Haze II atmospheric model containing the cloud and haze
aerosols described in Sec. \ref{Sec:cloudandhaze} are presented in
Fig.\ \ref{Fig:approxcomp}, where the calculation that ignores Raman
scattering is compared to the Raman calculation and to the modified Pollack and
tuned-$\omega$ approximations.  At UV-visible
wavelengths we see that the size of the Raman features is
substantially reduced by the addition of a high altitude absorbing
haze, which also depresses the baseline reflectivity as needed to
approximately match observations. The net effect of Raman scattering
on the baseline reflectivity is also reduced substantially compared to
its effect in a clear atmosphere. This is another example of the nonlinear
effects of adding absorption
(see discussion in Sec.\ \ref{Sec:kcorrect}). The
effects of the cloud layers are most easily seen at longer wavelengths where
the I/F in window regions and regions of intermediate methane absorption
are increased substantially compared to that computed for a clear atmosphere.

The tuned-$\omega$ approximation is seen to perform well for $\lambda
<$ 0.6 $\mu$m, where haze effects dominate, but is biased about 5\%
too high in the strongly absorbing regions at longer wavelengths,
although remaining quite accurate in window regions.  This is not too
surprising, because the tuning in these regions is involves a delicate
balance between large molecular absorption and large corrections of
that absorption to simulate the Raman-scattered contributions.  As
vertical structure is changed from that used for tuning, the tuning
can be easily upset.

The modified Pollack approximation performs somewhat worse at UV-visible wavelengths than
it did for the clear atmosphere case (Fig.\ \ref{Fig:approx3in1}). The overall negative
bias seen here could probably be largely eliminated by adjusting the parameters
of the $g_\ell$ factors in Eq.\ \ref{Eq:pollack}. That would do much to reduce the
positive errors in the window regions beyond 0.45 $\mu$m or the negative errors in the
absorbing regions.  These are inherent in the physical model on which the Pollack
approximation is based.  Overall, this approximation is less upset by the
presence of the cloud layers than is the tuned-$\omega$ approximation.

\begin{figure}[!htb]
\hspace{-0.2in}\includegraphics[width=3.7in]{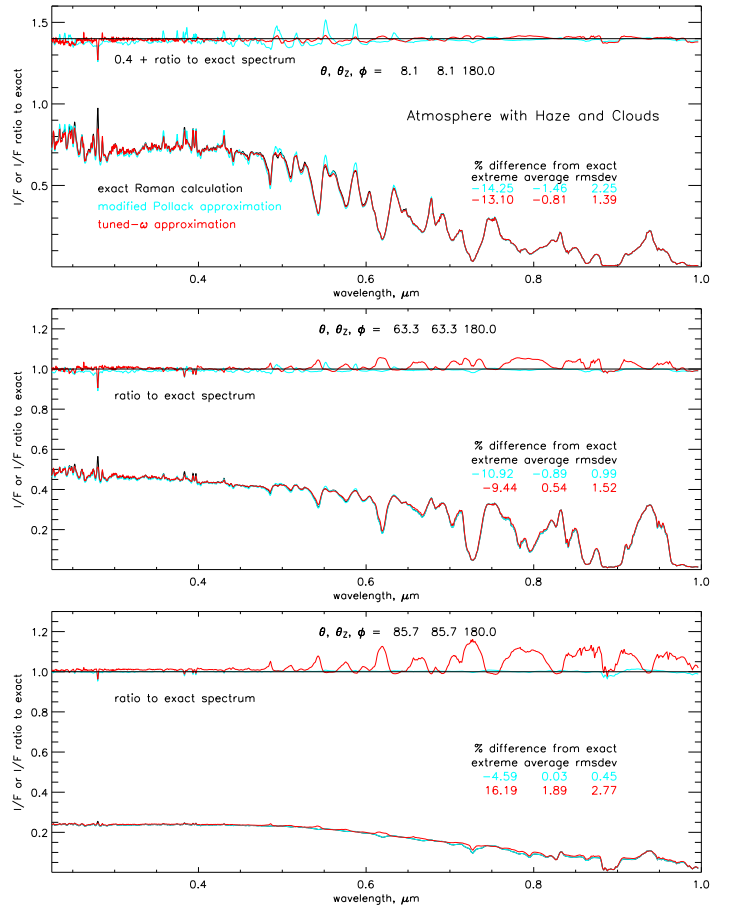}
\caption{Comparisons of Neptune's spectra at specific view angles for a sample 
aerosol structure containing an absorbing high-altitude haze and two
layers of clouds (Haze II model), using the modified Pollack
approximation (blue), the tuned-$\omega$ approximation (red), and the
exact method of calculation (black).  The ratio spectral are also
shown in each panel.}
\label{Fig:approxang}
\end{figure}

A comparison of these two approximations at three different view angles is presented
in Fig.\ \ref{Fig:approxang}.  We see that the tuned-$\omega$ approximation is
much better near the central disk, and effective at all wavelengths.  It is also effective
at all angles for wavelengths less than 0.47 $\mu$m.  But at longer wavelengths, errors
in the regions of strong methane absorption grow with increasing view angle, making
the modified Pollack approximation better for view angles beyond about 50\deg. 

The Karkoschka approximation was not included in this particular
comparison because is not useful for spatially resolved observations,
especially those with variable cloud structure, unless the model
coefficients are changed for each view angle and structure.  Even
then, the serious problems at $\lambda >$ 0.4 $\mu$m (see Fig.\
\ref{Fig:ramremove}) also reduce its utility, even for disk-average
analyses.

\section{Summary and Conclusions}

 Major results of this
 investigation can be summarized as follows:
\begin{description}
\setlength\itemsep{-0.02in}
\item[\textmd{(1) \it Radiation Transfer.}]  Procedures for modifying
the Evans and Stevens (1991) vector radiation transfer code to handle
Raman scattering are described and justified. Equations are
provided for the Raman differential generators. The use of a Raman
source matrix facilitates the computation of the first spatially
resolved Raman scattering spectra for Neptune.  Validation of
the code is obtained by demonstration of photon conservation,
comparison with prior low-resolution calculations, and comparison
with HST and groundbased observations.
 \setlength\itemsep{0.2in}

\item[\textmd{(2) \it Basic Effects on Geometric Albedo.}] Raman scattering reduces 
the baseline geometric albedo of a semi-infinite conservative Rayleigh
scattering atmosphere by about 25\% from the value of 0.791 it would
otherwise have.  In the presence of a high-altitude absorbing haze, the
effect of Raman scattering is reduced by about a factor of two. The
well known sharp positive spectral peaks from fill-in of absorption
lines in the solar spectrum can greatly exceed the Rayleigh
conservative limit. For photons introduced at 228 nm, the $Q$
transition source function for the first scattering has a broad peak
near 300 mb; after further scattering photons move deeper into the
atmosphere and leak out the top, moving the peak of the source
function further downward after each scattering. 
\setlength\itemsep{-0.02in}

\item[\textmd{(3) \it Near-IR Effects.}]  For a clear atmosphere about 4\% of the light in the deep
 methane bands in the near-IR is due to Raman scattering, while the
 near IR window regions are almost unaffected by Raman scattering.
 The conclusion of Cochran and Trafton (1978) that the light in the
 deep methane bands is almost entirely Raman scattered is a result of
 their assumption of what are now known to be excessively high methane
 mixing ratios in the stratosphere.

\item[\textmd{(4) \it Multiple Scattering.}] The calculated history of monochromatic incident
photons shows that multiple Raman scattering is quite significant in a
conservative Rayleigh scattering atmosphere, where the Raman
absorption process can compete with the loss to space.  In one example
calculation, only 40\% of the Raman scattered photons exiting the
atmosphere experienced only one Raman scattering.

\item[\textmd{(5) \it Angular Variation.}] At low phase angles spatially resolved
Raman spectra show that Raman spectral features are enhanced at near the center
of the planetary disk and suppressed at the limb. We also found that approximations
that work reasonably well in matching accurate disk-integrated calculations
do not do as well in matching observations in specific directions, especially
at large view angles.

\item[\textmd{(6) \it Comparison with Observations.}] Aerosol-free models of Neptune's
spectrum correlate well with observed spectral features, but confirm a
need for haze absorption to reduce the baseline geometric albedo and
to reduce the amplitude of spectral modulations induced by Raman
scattering.  IUE full-disk observations, adjusted to account for
current knowledge of of Neptune's size, have a baseline offset of
$\sim$6\% in the 0.24-0.26 $\mu$m range, increasing to about 10\% in
the 0.30-0.32 $\mu$m range, both of which are fit well by our Haze I
model, containing 0.2 optical depths (at 0.5 $\mu$m) of a UV absorbing
haze of 0.2-$\mu$m radius particles uniformly mixed between 0.1 and
0.8 bars.  FOS observations between 0.22 and 0.29 $\mu$m, have a
similar offset and high-resolution spectral structure that is also
well fit by our Haze I model.
Groundbased disk-integrated observations between 0.35 and
0.45 $\mu$m have a baseline offset that is 13\% below the
clear-atmosphere calculation which is well matched by our Haze I
calculation, although several of the spectral features are seen to
have greater amplitude in the calculations than in the
observations. We consider the Haze I model to be a sample calculation,
not a tightly constrained fit.

\item[\textmd{(7) \it Karkoschka correction.}]  The Karkoschka
(1994) method of applying Raman corrections to calculated spectra and
removing Raman effects from observed spectra is shown to be relatively
accurate under restricted conditions, but has practical limitations
for problems involving spatially resolved observations or for $\lambda >
.4$ $\mu$m.  His Raman removal algorithm undercorrects the depths of
weak CH$_4$ absorption bands, and thus corresponding absorption
coefficients derived from such spectra (Karkoschka 1994; 1998) will
need some revision. The relatively large amplitude of the Q-branch
contributions found by Karkoschka is shown to be consistent with
current estimates of Raman cross sections, not an indication that they
are in error.

\item[\textmd{(8) \it Wallace Approximation.}] We agree with Wallace (1972)
 that his suggested approximation yields geometric albedos that are
 typically $\sim$5\% lower than an accurate calculation. A
 generalization of this approximation that counts part of the
 vibrational cross section as contributing to scattering can remove
 this bias.  This approximation does not help reduce errors in the
 strong CH$_4$ bands and is mainly useful at low resolution and for
 $\lambda <$ 0.45 $\mu$m.

\item[\textmd{(9) \it Pollack Approximation.}] The initial form of the approximation suggested 
by Pollack \etal (1986) generates serious errors. But generalizing it to include
rotational and vibrational transitions, enforcing energy conservation,
and limiting the solar spectral ratio enhancement, can produce a relatively good
approximation, except for $\lambda >$ 0.45 $\mu$m, where ignoring Raman
scattering is just as accurate.

\item[\textmd{(10) \it Tuned-$\omega$ Approximation.}] By tuning the absorption attributed
to Raman scattering it is possible to make a non-Raman calculation
match a true Raman calculation with virtually no error.  If that
matching is done with respect to geometric albedo, it will not match
nearly as well for spatially resolved calculations. It is nevertheless
the best approximation to use under cloud-free conditions for zenith
angles less than 70\deg and wavelengths less than 0.85 $\mu$m. It is
the only one that provides improvements for $\lambda >$ 0.45
$\mu$m, although is not very effective for $\lambda >$ 0.47 $\mu$m when
cloud layers of intermediate opacity are present.
 
\end{description}
 
\noindent Most of these conclusions apply with equal force to Raman scattering in the
atmosphere of Uranus because of its similarity to Neptune in vertical
temperature structure, gas composition, and relatively low aerosol
loading.  Conclusion (6) would need the most revision to accommodate
differences in detailed observational results.

\section*{Acknowledgments}

This research was supported by a grant from NASA's Planetary
Atmospheres program and in part by an Archive Research Grant
from the Space Telescope Science Institute. I thank Pat Fry who helped
with the installation of the original Evans and Stevens code and
advised on FORTRAN debugging of modifications to it.  I also thank
Evans and Stevens for making their well written code available to the
community, R. Courtin for providing tabular data files of his processed
FOS observations, and two anonymous reviewers.

\section*{References}

\begin{description}

\item[] Allen, C.\ W. 1964. Astrophysical Quantities, 2nd Edition. 291 pages. Athlone
Press, London.
\setlength\itemsep{-0.02in}

\item[] Baines, K.\ H., and W. H. Smith 1990. The Atmospheric Structure and Dynamical
Properties of Neptune Derived from Ground-Based and IUE Spectrophotometry.
{\it Icarus \bf 85}, 65-108.

\item[] Baines, K.\ H., H.\ B. Hammel, K.\ A. Rages, P.\ N. Romani, and
R.\ E. Samuelson (1995) Clouds and hazes in the atmosphere of
Neptune. In\ {\it Neptune and Triton}, D.P. Cruikshank, ed., Univ. of
Arizona Press.

\item[] Baines, K.\ H., and H.\ B. Hammel 1994. Clouds, hazes, and the stratospheric
methane abundance in Neptune. {\it Icarus \bf 109}, 20-39.

\item[] B\'{e}tremieux, Y., and R.\ V. Yelle 1999. HST Detection of H$_2$ Raman
Scattering in the Jovian Atmosphere. {\it Icarus \bf 142}, 324-341.

\item[] Bhagavantam, S., 1931. Raman effect in gases I. Some experimental
results {\it Ind. J. Phys. \bf 6}, 319-330.

\item[] Bhagavantam, S., 1942. {\it Scattering of Light and the Raman Effect.},
333 pages, Chemical Publishing Company Inc., Brooklyn NY.

\item[] Cabbanes, J.  and A. Rousset 1936. Diffusion mol\'eculaire: Mesure
du facteur de d\'epolarization des raies Raman dans les gaz:
Azote, oxyg\`ene, gaz carbonique. {\it Comptes Rendus \bf 202}, 1825-1828.

\item[] Cochran, W. D., and L.\ M. Trafton 1978. Raman scattering in the atmospheres
of the major planets. {\it The Astrophys. J. \bf
219}, 756-762.

\item[] Conrath, B.J., T.C. Owen, and R.E. Samuelson 1993. Constraints
on N$_2$ in Neptune's atmosphere from Voyager measurements. {\it Icarus \bf
101}, 168-171.

\item[] Conrath, B.\ J., D. Gautier, G.\ F. Lindal, R.\ E. Samuelson, and W.\ A. Shaffer
1991. The helium abundance of Neptune from Voyager measurements. {\it
J. Geophys. Res. \bf 96}, 18907-18919.

\item[] Courtin, R. 1999. The Raman signature of H$_2$ in the UV spectra of Uranus
and Neptune: Constraints on the haze optical properties and on the para-H$_2$
fraction. {\it Planet. and Space Sci. \bf 47}, 1077-1100.

\item[] Davies,  M., V.\ K. Abalakin, A. Brahic, M. Bursa, B.\ H. Chovitz,
JK.\ H. Lieske, P.\ K. Seidelmann, A.\ T. Sinclair, and I.\ S. Tiuflin 1992.
Report of the IAU/IAG/COSPAR Working Group on Cartographic Coordinates
and Rotational Elements of the Planets and Satellites - 1991, {\it Celestial
Mechanics and Dynamical Astronomy (ISSN 0923-2958), \bf 53}, No. 4,
p. 377-397.

\item[] Dlugach, E.\ G., and E.\ G. Yanovitskij 1974. The Optical Properites of
 Venus and the Jovian Planets. II. Methods and Results of Calculations
 of the Intensity of Radiation Diffusely Reflected from Semi-infinite
 Homogeneous Atmospheres.  {\it Icarus \bf 22}, 66-81.

\item[] Evans, K.\ F., and G. L. Stephens 1991. A New Polarized Atmospheric
Radiative Transfer Model. {\it J. Quant. Spectr. and Rad. Transfer \bf
46}, 413-423.

\item[] Farkas, A., 1935.  Orthohydrogen, Parahydrogen and Heavy Hydrogen.
Cambridge Univ. Press, London.

\item[] Ford, A.\ L., and J.\ C. Browne 1973. Rayleigh and Raman Cross Sections
for the Hydrogen Molecule. {\it Atomic Data \bf 5}, 305-313.

\item[] French, R.\ G. 1984. Oblatenesses of Uranus and Neptune. In {\it Uranus and Neptune}, 
NASA Conf. Pub. 2330, 349-355.

\item[]	Gibbard, S. G., I. de Pater, H.\ G. Roe, S. Martin, B.\ A. Macintosh, and C.\ E. Max
2003. The altitude of Neptune cloud features from high-spatial-resolution near-infrared spectra.
{\it Icarus \bf 2}, 359-374. 

\item[] Goody, R.\ M., Y.\ L. Yung 1989. {\it Atmospheric Radiation}, Oxford
University Press, Oxford, New York.

\item[] Hansen, J.\ E. 1971. Multiple scattering of polarized light in 
planetary atmospheres. Part II. Sunlight reflected by terrestrial
water clouds. {\it J. Atmos. Sci. \bf 28}, 1400-1426.

\item[] Hansen, J.\ E., and L.\ D. Travis 1974. Light Scattering in Planetary
Atmospheres. {\it Space Science Reviews \bf 16}, 527-610.

\item[] Hinson, D.\ P., and J.\ A. Magalh\~aes 1993. Inertio-gravity waves in 
the atmosphere of Neptune. {\it Icarus \bf 99}, 142-161.

\item[] Hollas, M. 1992. {\it Modern Spectroscopy}, John Wiley \& Sons, New York.

\item[] Hovenier, J.\ W. 1969. Symmetry relationships for scattering of 
polarized light in a slab of randomly oriented particles. {\it J. Atmos. Sci. \bf 26},
488-494.

\item[] Hu, Y.-X., B. Wielicki, B. Lin, G. Gibson, S.l-C. Tsay, K. Stamnes, 
and T. Wong 2000. $\delta$-Fit: A fast and accurate treatment of particle
scattering phase functions with weighted singular-value decomposition
least-squares fitting.  {\it J. Quant. Spectr. and Rad. Transfer \bf
65} 681-690.

\item[] Karkoschka, E.  1994.  Spectrophotometry of the jovian
planets and Titan at 300- to 1000-nm wavelength: The methane
spectrum. {\it Icarus \bf 111}, 174-192.

\item[] Karksochka, E. 1998. Methane, ammonia, and temperature measurements of
the Jovian planets and Titan from CCD-spectroscopy. {\it Icarus \bf 133}, 134-146.

\item[] Karkoschka, E. and M. Tomasko 1992. Saturn's Upper Troposphere 1986-1989.
 {\it Icarus \bf 97}, 161-181.

\item[] Kattawar, G.\ W., and C.\ N. Adams 1971. Flux and Polarization
Reflected from a Rayleigh-scattering Planetary Atmosphere. {\it Astrophys.
J. \bf 167}, 183-192.

\item[] Kurucz, R.\ L.\ 1993. Smithsonian Astrophysical Obsevatory CD ROM No. 13.

\item[] Lindal, G.F. 1992. The Atmosphere of Neptune: An analysis of 
radio occultation data acquired with Voyager 2.  {\it Astron. J. \bf 103} ,
967-982.

\item[] Massie, S.\ T., and D.\ M. Hunten 1982. Conversion of para and ortho 
Hydrogen in the Jovian Planets. {\it Icarus \bf 49}, 213-226.

\item[] Mishchenko  Th, M.\ I., A.\ A. Lacis, and L.\ D. Travis  1994. Errors
induced by the neglect of polarization in radiance calculations for
Rayleigh-scattering atmospheres. {\it J. Quant. Spectr. and Rad. Transfer \bf
51}, 491-510.

\item[] Moses, J. I., K. Rages, and J. B. Pollack  1995. An Analysis of Neptune's
Stratospheric Haze Using High-Phase-Angle Voyager Images. {\it Icarus \bf 113},
232-266.

\item[] Neckel, H. and D. Labs  1984.  The solar radiation between 3300 and
12500 \AA .  {\it Solar Phys.  \bf 90}, 205-258.

\item[] Parthasarathy, S. 1951. Light scattering in gases. {\it Indian J. Phys. \bf 25}, 21-24.

\item[] Penndorf, R. 1957. Tables of the refractive index for standard air
and the Rayleigh scattering coefficient for the spectral region between
0.2 and 20 $\mu$m and their application to atmospheric optics. {\it J. Opt. Soc.
Am. \bf 47}, 176-182.

\item[] Placzek, G., 1959. {\it Rayleigh and Raman scattering}, pp. 206, 
University of California, Lawrence Radiation laboratory (Livermore, CA);
[translated from Handbuch der Radiologie, Ed. E. Marx, Vol. 6,
part II, end ed., (Akademische Verlagsgesellschaft, Leipzig, 1934), pp. 205-374]

\item[] Pollack, J.\ B., K. Rages, K.\ H. Baines, J.\ T. Bergstralh, D. Wenkert,
and G.\ E.\ Danielson 1986. Estimates of the Bolometric Albedos and
Radiation Balance of Uranus and Neptune. {\it Icarus \bf 65}, 442-466.

\item[] Pryor, W.R., R.A. West, K.E. Simmons, and M. Delitsky
1992. High-phase angle observations of Neptune at 2650 \AA and 7500\AA
: Haze structure and particle properties. {\it Icarus \bf 99},
302-316.

\item[] Rottman, G.\ J., T.\ N. Woods, T.\ P. Sparn 1993. SOLar-STellar Irradiance
Comparison Experiment I: 1. Instrument design and operation. {\it J. Geophys.
Res. \bf 98}, 10667-10677.

\item[] Savage, B.\ D., W. D. Cochran, and P. R. Wesselius 1980. Ultraviolet
Albedos of Uranus and Neptune {\it The Astrophys.
J. \bf 237}, 627-632.

\item[] Soris, C.\ E., and F.\ J. Evans  1999. Filling in of Fraunhofer and
gas-absorption lines in sky spectra as caused by rotational Raman
scattering. {\it Applied Optics \bf 38},
2706-2713.

\item[] Sromovsky, L.\ A., 2004. Effects of Rayleigh-scattering polarization
on reflected intensity: A fast and accurate approximation for atmospheres
with aerosols.  {\it Icarus \bf in press.} 

\item[] Sromovsky, L.\ A., P.\ M. Fry, S.\ S. Limaye, and K.\ H. Baines 2003.
The nature of Neptune's increasing brightness: evidence for a seasonal response.
{\it Icarus \bf 163}, 256-261.

\item[] Sromovsky, L.A., P.M. Fry, K.H. Baines, and T. Dowling 2001a.
Coordinated 1996 HST and IRTF Observations of Neptune
and Triton II: Implications of Disk-Integrated Photometry,
{\it Icarus \bf 149}, 435-458.

\item[] Sromovsky, L.A., P.M. Fry, T. Dowling, K.H. Baines, and S.S.
Limaye 2001b. Coordinated 1996 HST and IRTF Observations of Neptune
and Triton III: Neptune's Atmospheric Circulation and Cloud Structure,
{\it Icarus \bf 149}, 459-488.

\item[] Stamnes, K., S.-C. Tsay, W. Wiscombe, and K. Jayaweera 1988. 
Numerically stable algorithm for discrete-ordinate-method radiative
transfer in multiple scattering and emitting layered media. {\it
App. Opt. \bf 27}, 2502-2509.

\item[] Sweigart, A.\ V. 1970. Radiative Transfer in Atmospheres
scattering According to the Rayleigh Phase Function with Absorption.
{\it Astrophys. J. \bf 182}, 1-80. 

\item[] Toon, O.\ B., C.\ P. McKay, T.\ P. Ackerman, K. Santhanam 1989. Rapid
calculations of radiative transfer heating rates and photodissociation
rates in inhomogeneous multiple scattering atmospheres. {\it
J. Geophys. Res \bf 94}, 16287-16301.

\item[] Wagener, R., J. Caldwell, and K-H. Fricke 1986. The Geometric
Albedos of Uranus and Neptune between 2100 and 3350 \AA. {\it Icarus \bf 67},
281-288.

\item[] Wallace, L., 1972. Rayleigh and Raman Scattering by H$_2$ in a Planetary
Atmosphere. {\it Astrophys. J. \bf 176}, 249-257.

\item[] Woods, T.\ N., G.\ J. Ucker, G.\ J. Rottman 1993.  SOLar-STellar Irradiance
Comparison Experiment I: 2. Instrument calibration. {\it J. Geophys.
Res. \bf 98}, 10679-10694.

\end{description}

\end{document}